\documentclass[final,3p,times]{elsarticle}

\usepackage{graphics}
\usepackage{amsmath}
\usepackage{latexsym}

\journal{Journal of Computational and Applied Mathematics}

\usepackage{color}
\usepackage{times}
\usepackage{url}
\usepackage{array}
\usepackage{stmaryrd}
\usepackage{psfrag}
\usepackage{rotating}
\usepackage{placeins}

\renewcommand{\vec}[1]{{\bf {#1}}}
\newcommand{\Ti}{{\mathcal{T}_i}}
\newcommand{\Th}{{\mathcal{T}_h}}

\newcommand{\dd}{\text{d}}
\newcommand{\llbrace}{\{\hspace{-1mm}\{}
\newcommand{\rrbrace}{\}\hspace{-1mm}\}}

\newcommand{\vE}{\vec{E}}
\newcommand{\vx}{\vec{x}}
\newcommand{\vxn}{\hat{\vx}}
\newcommand{\vy}{\vec{y}}

\begin{document}

\begin{frontmatter}
%
%
\title{Error-Driven Dynamical $hp$-Meshes with the Discontinuous Galerkin Method for Three-Dimensional Wave Propagation Problems}
\author[ifh]{Sascha M. Schnepp}\ead{schnepps@ethz.ch}
\fntext[label2]{The author acknowledges the support of the `Alexander von Humboldt-Foundation' through a `Feodor Lynen Research Fellowship'.}
\address[ifh]{Laboratory for Electromagnetic Fields and Microwave Electronics, ETH Zurich, Gloriastrasse 35,  8092 Zurich, Switzerland}
%
%
%
%

\begin{abstract}
An $hp$-adaptive Discontinuous Galerkin Method for electromagnetic wave
propagation phenomena in the time-domain is proposed. 
The method is highly efficient and allows for the first time the adaptive full-wave simulation of large, time-dependent problems in
three-dimensional space. 
Refinement is performed anisotropically in the approximation order $p$
and the mesh step size $h$ regardless of the resulting level of
hanging nodes. For guiding the adaptation process a variant of the
concept of reference solutions with largely reduced computational costs is proposed.
The computational mesh is adapted such that a given error tolerance is respected throughout
the entire time-domain simulation.
\end{abstract}


\end{frontmatter}

\section{Introduction}
\label{sec:introduction}

In this article, we are concerned with adaptively solving the Maxwell equations
for electromagnetic fields with arbitrary time dependence
in a three-dimensional domain such that a prescribed error tolerance is respected. 
In order to achieve this goal the Discontinuous Galerkin Method (DGM)~\cite{Reed1973,LeSaintRaviart1974} is
applied on anisotropic $hp$-meshes,
which dynamically and autonomously adapt as the electromagnetic
fields evolve. The mesh refinement is driven by a robust local error estimate based on
a modification of the so-called method of reference solutions~\cite{Demkowicz1,Solin2008117}
with largely reduced numerical costs. 

The DG method has gained wide acceptance
as a high order numerical method, which is very well-suited for time-domain problems.
It combines the usually opposing key features of high order accuracy and flexibility. 
In particular, the method can easily deal with meshes containing hanging nodes, which makes it particularly well suited for $hp$-adaptivity.
There is a well established body of literature on the DG method for
various types of problems available. It has been thoroughly investigated by several research
groups (see~e.g.~\cite{Cockburn2001,Hesthaven2008,Fezoui2005} and
references therein). Concerning Maxwell's equations in time-domain, the DGM has been studied in particular
in~\cite{Fezoui2005,Hesthaven2002Maxwell,Cohen2006,Gjonaj2006}. 
The latter two make use of hexahedral meshes, which allow for a computationally more efficient
implementation~\cite{Wirasaet:2010p1362}.

The simplest approach to adapted grids consists of static \textit{a priori} $h$-refinement
around edges and corners, i.e., the possible locations of field singularities~\cite{Fahs2007}.
While this approach mitigates negative effects of fields singularities on the global solution accuracy, 
the level of refinement to be applied for achieving a certain accuracy is unknown. Moreover,
edges and corners require no mesh refinement while there is no field, for instance, before illumination by a wave
or after scattering took place. It also remains unclear how to choose polynomial orders in the remaining
mesh. For these reasons our focus is on $hp$-adaptivity based on error estimations of the 
time-dependent solution.

Mesh refinement and specifically $hp$-adaptation has received
considerable and continuous attention.
The first published work on $h$-, $p$- and $hp$-adaptivity
within the DG framework is presumably \cite{Bey:1995kj},
where the authors considered linear scalar hyperbolic conservation laws in two-dimensional
space. Hyperbolic problems have also been addressed, e.g., by
Flaherty, Shephard and co-workers who considered two-dimensional problems in~\cite{Devine:1996ju,Remacle2003}
as well as three-dimensional settings with pure $h$-refinement in~\cite{Flaherty:1997de,Remacle:2005ef}.
A large number of contributions has been authored by Houston and various co-workers.
They present a number of approaches to adaptivity and deal with first-order
hyperbolic problems in \cite{Houston2001,Houston:2002p149}, using 
adjoint solutions \cite{Houston:2002p149,Hartmann:2003gl} or estimating
errors in an energy norm \cite{Houston:2005p133,Houston:2007dp}. The contributions have a clear focus on the
rigorous derivation of error estimates and error bounds. Applications are limited
to one or two space dimensions.
Recently, Solin and co-workers published papers, where they apply dynamical $hp$-meshes
for various coupled problems including electromagnetics in two space dimensions 
\cite{Solin:2010tm,Dubcova:2010kea,Korous:2012wi}.
They employ the concept of reference solutions for controlling mesh adaptivity and perform
refinements, which are fully anisotropic in both mesh parameters $h$ and $p$.
The application of reference solutions in their original form is numerically very expensive.
In \cite{Korous:2012wi} it is stated that the solution of large three-dimensional
problems would require distributed parallel computing.



In this paper, we propose a modification of the concept of reference solutions with drastically reduced numerical costs,
which makes such simulations feasible. At the same time the key advantages are maintained,
in particular its robustness and the independence of a particular set of underlying partial 
differential equations. The increased efficiency comes
at the price of losing some sharpness in the error estimate.
Like the original formulation,
the proposed algorithm is entirely devoid of tuning parameters, 
and it reduces the true approximation error, i.e., it is not based on
residuals or heuristic measures such as steep gradients.
The adaptation can be performed in four major modes: isotropic in $h$ and $p$, 
anisotropic in one of $h$ or $p$, and fully anisotropic in $h$ and $p$. Unconstrained
refinement in $h$ is possible because we allow for high level hanging nodes. 
The number of degrees of freedom (DoF) in a discretization will 
usually decrease from the former to the latter mode,
while the computational load for finding the adapted mesh increases. However, we will
show below that great savings in both, the number of DoF and computational time can be
achieved by using fully anisotropic adaptivity.




The remainder of this article is organized as follows.
In Sec.~\ref{sec:discr-maxw-equat} the notation and
Finite Element Spaces (FES) are introduced, which are applied
for obtaining a weak DG formulation of Maxwell's equations.
Section~\ref{sec:autom-dynam-mesh} is devoted to the 
mesh refinement algorithm. First the individual steps, which
constitute an adaptive algorithm are discussed. They are
error estimation, element marking, the $h$--$p$-decision and the
actual mesh adaptation. For each step a brief description with a 
review of the state of the art is provided, before we proceed with
the details of our realization of each step in the Sections~\ref{sec:error-estimation} to \ref{sec:comm-pract-issu}.
Examples are presented in Sec.~\ref{sec:examples}, which include
a waveguide and an antenna radiation problem.
Section~\ref{sec:conclusion} summarizes the findings and concludes
the article.

\section{Discretization of Maxwell's Equations}
\label{sec:discr-maxw-equat}

In the following we assume resting, heterogeneous, linear, isotropic,
non-dispersive and time-independent materials. Then, the magnetic permeability, $\mu$, and dielectric
permittivity, $\epsilon$, are scalar values depending on
the spatial position only. Under these assumptions Maxwell's equations read
\begin{eqnarray}
  \label{eq:Faraday}
  \nabla \times \vec{E} (\vec{x},t) &=& - \mu(\vec{x}) \frac{\partial}{\partial t} \vec{H}(\vec{x},t), \\
  \label{eq:Ampere}
  \nabla \times \vec{H} (\vec{x},t) &=& \textcolor{white}{-}\epsilon(\vec{x})\frac{\partial}{\partial t} \vec{E} (\vec{x},t) + \vec{J}(\vec{x},t),
\end{eqnarray}
with the spatial variable $\vec{x} \in \Omega \subset \mathbb{R}^3$ and the temporal variable $t \in [t_0,T] \subset \mathbb{R}$
subject to boundary conditions specified at the domain boundary $\partial\Omega$ and initial conditions specified at time $t_0$.
The electric and magnetic field vectors are denoted by $\vec{E}$ and $\vec{H}$,
$\vec{J}$ denotes the electric current density. 

Discretizations of Maxwell's equations using the Discontinuous Galerkin Method have  been
obtained among others in~\cite{Fezoui2005,Hesthaven2002Maxwell,Cohen2006,Gjonaj2006}.
We will follow the framework and notation described in our previous work~\cite{Schnepp:2012JCAM},
which makes use of hexahedral meshes and modal basis functions as introduced in~\cite{Gjonaj2006}. 

\subsection{Notation}
\label{sec:notation}

We denote by $\mathcal{T}_h$ a tessellation of the domain of interest $\Omega$ composed from
non-overlapping hexahedra $\mathcal{T}_i$ such that $\mathcal{T}_h = \bigcup_{i=1}^N \mathcal{T}_i$
covers $\Omega$. The tessellation is required to be derivable from a regular root tessellation $\mathcal{T}_0$
by means of element bisections. However, we do not demand the resulting tessellation
to be regular, i.e., we allow for hanging nodes and specifically for high level hanging
nodes. The number of bisections performed for obtaining element $\mathcal{T}_i$
is denoted by $L_i$ in the isotropic and $L_{d,i}$ in the anisotropic case where $d$ corresponds
to any of the spatial coordinates $\{x,y,z\}$.
We call the intersection of two neighboring elements $\mathcal{T}_i \cap \mathcal{T}_k$ their interface $\mathcal{I}_{ik}$. 
In non-conformingly refined meshes, every face $\mathcal{F}_j$ of a hexahedral element may
be partitioned into several interfaces depending on the number of neighbors $K$ such that
$\mathcal{F}_j = \bigcup_{k=1}^K \mathcal{I}_{ik}$. This is an important difference to most other
works including~\cite{Fezoui2005,Cohen2006,Hesthaven:2004jt}, which require
one-to-one neighborhood relations. The (inter-)face orientation
is described by the outward pointing unitary normal $\mathbf{n}_j$. 
The union of all faces is denoted by $\mathcal{F}$. The volume
and edge length measures of element $i$ are denoted by $|\mathcal{T}_i|$ and $|\mathcal{T}_{d,i}|$.

\subsection{Finite Element Spaces and Approximations}
\label{sec:finite-elem-spac}

In DG methods trial and test functions are defined with element-wise compact support
\begin{equation}
  \label{eq:basisFcts}
  \varphi_i^p ({\bf x}) =
  \begin{cases}
    \varphi^p({\bf x}), &\mathbf{x} \in \mathcal{T}_i,\\
    0,                                & \text{otherwise}.
  \end{cases}
\end{equation}
Cartesian grids allow the application of tensor product basis functions of the form
\begin{equation}
  \label{eq:tensorBasis}
  \varphi^p(\vec{x}) = \bigotimes_{d\, \in \,\{x,y,z\}}\varphi^{p_d}(d),
\end{equation}
where $p$ is a multi-index obtained from all $p_d = 0..P_d$.
The local finite element spaces $\mathcal{V}^P(\Ti)$ spanned by the basis functions
are given by the tensor product of the respective one-dimensional spaces
\begin{align}
  \label{eq:DGtensorSpace}
  &(\mathcal{V}^P)_\Ti = (\mathcal{V}_x^{P_x})_{\mathcal{T}_{x,i}} \otimes (\mathcal{V}_y^{P_y})_{\mathcal{T}_{y,i}} \otimes (\mathcal{V}_z^{P_z})_{\mathcal{T}_{z,i}}, \hspace{1mm} \text{where} \\
  &(\mathcal{V}_d^{P_d})_{\mathcal{T}_{d,i}} = \text{span}\{ \varphi_i^{p_d }(d);\, 0 \le p_d \le P_d \}.
\end{align}
The approximation may, thus,
make use of different orders $P_d$ in each of the coordinate directions, where
the subscript is dropped if they are equal. We do not choose an interpolatory basis but
follow a spectral approach and apply Legendre polynomials
scaled such that~\cite{Gjonaj2006}
\begin{equation}
  \label{eq:orthogonalBasis}
  \int_{\mathcal{T}_{d,i}} \varphi^{p_d}_i(x)\varphi^{q_d}_i(x)\, \dd x = 
\begin{cases}
|\mathcal{T}_{d,i}|, \quad &p_d = q_d\\
0, \quad &\text{otherwise}.
\end{cases}
\end{equation}

Associating an FES~\eqref{eq:DGtensorSpace} with each element $\mathcal{T}_i$
of the tesselation defines the Finite Element discretization, where the
electric and magnetic field approximations ${\bf E}_h$ and ${\bf H}_h$ are represented as
\begin{equation}
  \label{eq:eglobal}
      {\bf E}({\bf x},t) \approx  {\bf E}_h ({\bf x},t) =
    \bigoplus_{i = 1}^N {\bf E}_i({\bf x},t), \quad \quad
      {\bf H}({\bf x},t) \approx  {\bf H}_h ({\bf x},t) =
    \bigoplus_{i = 1}^N {\bf H}_i({\bf x},t),
\end{equation}
with the element local representations
\begin{equation}
  \label{eq:elocal}
      {\bf E}_i({\bf x},t) = \sum_{p} {\bf e}^{p}_i(t) \varphi^{p}_i({\bf x}), \quad \quad
      {\bf H}_i({\bf x},t) = \sum_{p} {\bf h}^{p}_i(t) \varphi^{p}_i({\bf x}).
\end{equation}
The time-dependent vectors of coefficients $\vec{e} = (\vec{e}_1^0,..,\vec{e}_1^P,..,\vec{e}_N^0,..,\vec{e}_N^P)^\text{T}$
and $\vec{h} = (\vec{h}_1^0,..,\vec{h}_1^P,..,\vec{h}_N^0,..,\vec{h}_N^P)^\text{T}$ are the numerical degrees of freedom.

\subsection{Weak DG formulation}
\label{sec:weak-dg-formulation}

Following the
Galerkin procedure \eqref{eq:Faraday} and \eqref{eq:Ampere} are multiplied by a test function $\psi$
and integrated over the domain $\Omega$. Due to the compact support property~\eqref{eq:basisFcts}
the integration can be carried out over every element $\mathcal{T}_i$ individually. Next,
we perform integration by parts of the curl-terms and replace the exact field solution with the
approximations~\eqref{eq:eglobal}. This leads to the semi-discrete variational problem of finding
$\vec{e}$ and $\vec{h}$ such that
\begin{eqnarray}
  \label{eq:DGFaraday}
  \int_{\Ti} \psi \, \mu \frac{\partial}{\partial t}{\bf H}_h \:\dd^3 {\bf x} 
  - \int_{\Ti}(\mathbf{\nabla} \psi) \times {\bf E}_h \:\dd^3 {\bf x} 
  +\int_{\partial \Ti} \psi \, (\mathbf{n} \times \widehat{\bf E}_h)  \, \dd^2 {\bf x}  &=& 0\\
  \label{eq:DGAmpere}
  \int_{\Ti}  \psi \, \epsilon \frac{\partial}{\partial t}{\bf E}_h \:\dd^3 {\bf x} 
    + \int_{\Ti} (\mathbf{\nabla} \psi )\times {\bf H}_h \:\dd^3 {\bf x} 
  -\int_{\partial \Ti} \psi \, (\mathbf{n} \times\widehat{\bf H}_h)  \: \dd^2 {\bf x}  &=& 0,
\end{eqnarray}
$\forall i = 1,..,N;\, \forall \psi \in \mathcal{V}_i$. For the above equations
to be well-defined it is required that $\psi \in H^1$ in the interior of $\Ti$, which is 
fulfilled for the chosen Legendre basis.
Note that $\widehat{\bf E}_h$ and $\widehat{\bf H}_h$ denote
the numerical trace of the electric and magnetic field, which is single-valued for each
vector field component at element boundaries.
Introducing the numerical trace is a necessary step for resolving the ambiguity of the numerical
approximations~\eqref{eq:eglobal} at element interfaces. Due to the definition
of the basis function support in~\eqref{eq:basisFcts}, the components for the vector fields
$\vec{E}_h$ and $\vec{H}_h$ are single valued
at all points $\vec{x} \in \mathcal{T} \backslash \mathcal{F}$ but double-valued for all $\vec{x} \in \mathcal{F}$.
The numerical trace is computed as
\begin{equation}
  \label{eq:upwE}
    \widehat{\vec{E}}_{ik} = \llbrace \vec{E}\rrbrace_{\mathcal{I}_{ik}} 
    + \gamma \frac{\vec{n}_{ik} \times \llbracket \vec{H}\rrbracket_{\mathcal{I}_{ik}}}{2\llbrace Y \rrbrace_{\mathcal{I}_{ik}}}, \quad \quad
    \widehat{\vec{H}}_{ik} = \llbrace \vec{H}\rrbrace_{\mathcal{I}_{ik}}
    - \gamma \frac{\vec{n}_{ik} \times \llbracket \vec{E}\rrbracket_{\mathcal{I}_{ik}}}{2\llbrace Z \rrbrace_{\mathcal{I}_{ik}}}.
\end{equation}
Typical choices are the centered and upwind value obtained by setting $\gamma$ to zero or one, respectively,
where the upwind value is the solution of the Riemannian problem~\cite{LeVeque1990}.
Above $\llbrace \cdot \rrbrace$ and $\llbracket \cdot \rrbracket$ denote the average and jump operators
\begin{equation}
  \label{eq:avg}
  \llbrace \vec{a} \rrbrace_{\mathcal{I}_{ik}} = (  \vec{a}_{k|\mathcal{I}_{ik}} + \vec{a}_{i|\mathcal{I}_{ik}} )/2, \quad \quad
  \llbracket \vec{a} \rrbracket_{\mathcal{I}_{ik}} = \vec{a}_{k|\mathcal{I}_{ik}} - \vec{a}_{i|\mathcal{I}_{ik}}.
\end{equation}
The intrinsic impedance and admittance are given as
\begin{equation}
  \label{eq:FVZandY}
  Z = \sqrt{\frac{\epsilon}{\mu}}, \quad Y = \frac{1}{Z}.
\end{equation}
The surface integrals in~\eqref{eq:DGFaraday} and \eqref{eq:DGAmpere} represent interelement fluxes, 
the volume integrals are referred to as the mass and stiffness terms according to
standard FE nomenclature. In the following the dependence of the spatial and temporal variable
is not written down explicitly.

Note that no assumptions on the grid regularity have
been made in the derivation. This is in a sharp contrast with Finite Element Methods
based on edge elements, which require augmentation by edge constraints if hanging nodes
are to be included~\cite{Solin2008117}. In DG-type methods non-regular grids are no
methodological issue, they only make the implementation more involved.
The relative ease of handling non-regular
meshes combined with the strictly element-local character of the numerical
approximation make DG methods an ideal candidate for $hp$-adaptivity.

\section{Automatic and dynamic $hp$-adaptation}
\label{sec:autom-dynam-mesh}

Devising an $hp$-adaptive algorithm requires four major steps.
\begin{enumerate}
\item Derivation of global and local error estimates
\item Definition of a marking strategy for assigning a refinement/derefinement label to each element
\item Deriving criteria for making the $h$--$p$-decision
\item Definition of the actual mesh refinement/derefinement operators
\end{enumerate}

For each of these steps several alternatives are possible. We will briefly list a few popular techniques
and describe the main underlying idea before describing the approach followed in this
contribution along with the reasoning behind this choice.

\subsection{Error Estimation}
\label{sec:error-estimation}

Error estimators or indicators can, for instance, be obtained by expressing a
residual through the numerical approximation. Residual based
estimators in the context of Maxwell's equations
have been developed, e.g., in \cite{Houston:2005p133,Schnepp:2012JCAM,Beck:2002hm}, 
or in~\cite{Cockburn2003a,Barth2005} with applications outside electrodynamics.
Highly accurate estimators can be constructed based on adjoint solutions~\cite{Giles:2003eh,Wang:2009kv},
where the latter one is applied in a DG setting. However, the accuracy of adjoint
based estimators comes at the price of having to repeatedly solve for the adjoint problem in addition.
Comprehensive overviews of error estimation techniques are found in~\cite{VerfurthReview,Ainsworth:2000tw,Barth:2005ff}.

In this article we employ the concept of 
reference solutions~\cite{Demkowicz1,Solin2008117,Solin:2010p1978}
for obtaining error estimates. A reference solution is a numerically
computed approximation, which is assumed to be significantly more accurate
than the present approximation. This can be achieved by performing
one uniform $h$-refinement step combined with increasing the
approximation order by one in the element under consideration. 
Obtaining a reference mesh by pure $p$-enrichment has been
proposed as well. Both techniques provide a reference
mesh based on hierarchic FES enrichment.
We apply the concept in its
original form for finding an initial $hp$-mesh and propose
a modified, computationally much cheaper variant, which
is applied during the transient analysis.
We found this estimator to be very robust and find reliable estimates
independent of the local solution smoothness. This is an important advantage
over the residual based estimate proposed in~\cite{Schnepp:2012JCAM}.

The aim then is to find the minimal $hp$-mesh such that
\begin{equation}
  \label{eq:aimToleranceExact}
  \| \vec{\varepsilon} \|_{\mathcal{T}_h} = \| \vec{u} - \vec{u}_{hp}\|_{\mathcal{T}_h} = \bigg(\sum_{\Ti \in \mathcal{T}} \| \vec{u} - \vec{u}_{hp}\|_{\Ti}^2\bigg)^\frac{1}{2} \le \mathtt{TOL},
\end{equation}
where $\|\cdot\|_{\mathcal{T}_h}$ denotes the global $L^2$-norm, and it is taken into account that the solution is a
vector field. In the following $\vec{u}$ is used for denoting the electromagnetic solution
$(\vec{E},\vec{H})$. As only the approximation $\vec{u}_{hp}$ is known 
but not the exact solution $\vec{u}$ the target \eqref{eq:aimToleranceExact}
cannot be achieved directly. However, it can be achieved
asymptotically as
\begin{equation}
  \label{eq:aimToleranceApprox}
  \| \vec{\varepsilon}_{hp} \|_{\mathcal{T}_h} = \bigg(\sum_{\Ti \in \mathcal{T}}\| \vec{u}_\text{ref} - \Pi_{hp}\vec{u}_\text{ref}\|_\Ti^2 \bigg)^\frac{1}{2} 
  = \bigg(\sum_{\Ti \in \mathcal{T}}\| \varepsilon_{hp} \|_\Ti^2 \bigg)^\frac{1}{2},
\end{equation}
where $\Pi_{hp}$ is a projection operator from the enriched reference FES $\mathcal{V}_\text{ref}$ to
a space $\mathcal{V}_\text{c}$ associated with a refinement candidate.
The space $\mathcal{V}_\text{c}$
is reduced with respect to the reference space but enriched with respect to $\mathcal{V}_\Ti$ such that
\begin{equation}
  \label{eq:spaceNesting}
  \mathcal{V}_\Ti \subset \mathcal{V}_\text{c} \subset \mathcal{V}_\text{ref}. 
\end{equation}
Refinement can be anisotropic 
in one or both of the mesh parameters. From~\eqref{eq:aimToleranceApprox} it follows
that the element-wise error estimate is given as
\begin{equation}
  \label{eq:elementErrorEstimate}
  \| \varepsilon_{hp} \|_\Ti = \| \vec{u}_\text{ref} - \Pi_{hp}\vec{u}_\text{ref}\|_\Ti.
\end{equation}
Given a reference solution, the global and local error
estimates~\eqref{eq:aimToleranceApprox} and~\eqref{eq:elementErrorEstimate}
are fully computable.


\subsubsection{Initial Mesh}
\label{sec:initial-mesh}

Starting from the root tesselation $\mathcal{T}_0$ with 
some uniform polynomial order $P_0$ a reference mesh is constructed
by performing one uniform refinement step in $h$ and $p$.
We note that this has not to be done globally, but it can rather be done
consecutively with each element of the current tesselation.
The DG approximation $f_\Ti$ of a given function $f$ on the element $\Ti$,
is obtained by applying the orthogonal projection operator $\Pi$
\begin{equation}
  \label{eq:projectionOperator}
  f_\Ti = (\Pi f)_\Ti =  \sum_p (\Pi^p f)_{\mathcal{T}_i} \,  \varphi_i^p =
  \sum_p\frac{\left(\varphi_i^p, f \right)_{\mathcal{T}_i} }{\left(\varphi_i^p, \varphi_i^p \right)_{\mathcal{T}_i}}
  \varphi_i^p,
\end{equation}
where $(u,v)_{\mathcal{T}_i}$ denotes the inner product $\int_{\mathcal{T}_i} uv\, \text{d} \vec{x}$ on the element $\mathcal{T}_i$.
Equipping the FES~\eqref{eq:DGtensorSpace} with an inner product
defines a Hilbert space. Hence, the above
projector yields the best approximation in the $L^2$-sense.
After projecting the initial data to the refined elements
the approximation error $\varepsilon_i$ of element $\Ti$ is estimated
using~\eqref{eq:elementErrorEstimate}. This procedure is repeated
for all elements of the current tesselation. The global error $\varepsilon_{hp}$ 
is obtained from~\eqref{eq:aimToleranceApprox}. The construction of the initial $hp$-mesh
terminates when the stopping criterion $\varepsilon_{hp} \le \mathtt{TOL}$ is met.

\subsubsection{Dynamical Mesh}
\label{sec:dynamical-mesh}

In the construction of an optimal initial $hp$-mesh the reference solution
at each iteration can be generated because the initial data is known exactly.
Obviously, this approach cannot be transferred immediately to the transient analysis.
In~\cite{Solin:2010tm} the authors approach the transient case by employing
Rothe's method. In contrast to the widely used Method of Lines, Rothe's method
discretizes the time variable first while preserving continuity of the spatial
variable. This approach allows for applying the same techniques in the transient analysis that
were used for obtaining the initial mesh at the cost of having to solve for a system
of equations in every time step.

For performance reasons we prefer to employ explicit time-integration.
The straightforward extension in this case is to compute on two meshes,
the $hp$-mesh fulfilling the error tolerance and its reference mesh. However,
this approach can easily become prohibitively expensive, both in computing time and memory
consumption as the reference mesh usually has about 15 to 60 times more DoF
depending on the approximation order. 
Taking into account that, moreover, the reference solution
largely exceeds the required accuracy and is employed for driving the
adaptivity only, this solution does not appear to be ideal.
This motivated us to seek a different approach.

To this end, we switch roles of the reference mesh and the mesh
used for estimating the local error and claim that the approximation
on the current $hp$-mesh is sufficiently accurate for serving as 
the reference solution. Then, we estimate the element error by comparing
to a reduced FES. This FES can be obtained by derefining the mesh in
$h$ and $p$, or in a significantly more efficient
manner by reducing the approximation order $P$. 
The element-wise error estimate is computed as
\begin{equation}
  \label{eq:errorEstimateDynamicLocal}
  \|\varepsilon_{hp}\|_\Ti = \| \vec{u}_\text{ref} - \Pi_{p}\vec{u}_\text{ref}\|_\Ti.
\end{equation}
Here, the solution on the current $hp$-mesh is the reference solution and $\Pi_p$
is the projection operator to the $p$-reduced FES.
Computing the estimate~\eqref{eq:errorEstimateDynamicLocal} is very cheap.
As the basis is hierarchic it comes down to considering the highest order terms of the current
approximation only
\begin{equation}
  \label{eq:errorEstimateDynamicLocal2}
  \vec{u}_{\text{ref},\Ti} - \Pi_{p}\vec{u}_{\text{ref} ,\Ti}
  = \sum_{p = 0}^P \vec{u}^p_i \varphi^p_i - \sum_{p = 0}^{P-1} \vec{u}^p_i \varphi^p_i,
\end{equation}
where $\vec{u}_i^p$ denotes the vector of coefficients of order $p$ local to element $\Ti$.
Recalling that $p$ and $P$ are multi-indices as defined in~\eqref{eq:tensorBasis},
the local error estimate \eqref{eq:errorEstimateDynamicLocal2} is computed as
\begin{equation}
  \label{eq:errorEstimateDynamicLocal3}
  \|\varepsilon_{hp}\|_\Ti = \| \Big(\sum_{p_y=0}^{P_y}\sum_{p_z=0}^{P_z} \vec{u}_i^{p_xp_yp_z} \varphi_i^{p_xp_yp_z}\Big)_{|p_x=P_x} 
  +\: \Big(\sum_{p_x=0}^{P_x-1}\sum_{p_z=0}^{P_z} \vec{u}_i^{p_xp_yp_z} \varphi_i^{p_xp_yp_z}\Big)_{|p_y=P_y} 
  +\: \Big(\sum_{p_x=0}^{P_x-1}\sum_{p_y=0}^{P_y-1} \vec{u}_i^{p_xp_yp_z} \varphi_i^{p_xp_yp_z}\Big)_{|p_z=P_z} \|_\Ti.
\end{equation}
Evaluating the $L^2$-norm and inserting the scaling property~\eqref{eq:orthogonalBasis} 
yields the following form of the estimate
\begin{equation}
  \label{eq:errorEstimateDynamicLocal4}
  \|\varepsilon_{hp}\|_\Ti = \Big[ \Big( 
             \big( \sum_{p_y=0}^{P_y}\sum_{p_z=0}^{P_z} \|\vec{u}_i^{p_xp_yp_z}\|_2^2 \big)_{|p_x=P_x}
  +\: \big(\sum_{p_x=0}^{P_x-1}\sum_{p_z=0}^{P_z} \|\vec{u}_i^{p_xp_yp_z}\|_2^2\big)_{|p_y=P_y} 
  +\: \big(\sum_{p_x=0}^{P_x-1}\sum_{p_y=0}^{P_y-1} \|\vec{u}_i^{p_xp_yp_z}\|_2^2\big)_{|p_z=P_z} \Big)|\Ti| \Big]^{1/2}.
\end{equation}
It is an important feature that the computation of this estimate is highly efficient as 
no runtime quadratures have to be performed. 

We admit that the
approach of projecting the solution to a reduced FES instead
of an enriched one negatively affects the accuracy of the error estimation. 
However, it drastically reduces computational costs rendering the method
applicable for a much larger class of real world problems. In Sec.~\ref{sec:examples}
we demonstrate the robustness and reliability of this approach.

\subsection{Marking Strategy}
\label{sec:marking-strategy}

Following the error estimation, each element is
assigned one of the labels \textit{refine}, \textit{derefine} or \textit{retain} according to the marking strategy.
The marking strategy, hence, has a strong impact on the number of DoF in the computational mesh.
Popular strategies include error equidistribution, the fixed fraction strategy or variable
fraction strategies such as bulk-chasing, commonly known as D\"{o}rfler-marking.
The goal of the former strategy is to equilibrate the local errors by refining
or derefining elements such that $\varepsilon_i \approx \mathtt{TOL}/\sqrt{N}$,
where $\varepsilon_i$ is the local error estimate and $\mathtt{TOL}$ is a
user-defined error tolerance~\cite{Chen:1994hw}.
For the fixed and variable fraction strategies, the elements are ordered by
their estimated error at each refinement step. Then, for the former approach,
a fixed fraction of elements from the top and bottom are marked for refinement
and derefinement. The variable fraction or D\"{o}rfler-marking on the other hand
continues to mark elements from the top and bottom of the list until their accumulated
error accounts for a certain percentage of the total error. This can be expressed as
finding a minimal subset ${\mathcal{T}}^+_h$ and a maximal subset ${\mathcal{T}}_h^-$ of $\mathcal{T}_h$ such that
$\sum_{\Ti \in \mathcal{T}_h^{\{+,-\}}} \varepsilon^2_i \ge \theta^2_{\{+,-\}} \sum_{{\mathcal{T}_i} \in {\mathcal{T}}_h} \varepsilon^2_i$,
where the sign indicates refinement and derefinement.
As the values of $\theta_{\{+,-\}}$ indicate fractions of the total error, the
D\"{o}rfler-marking can be considered as a fixed fraction marking with respect to
the total error.
Often a few percent of the elements make up for more than 90 \% of the total error,
while most of the elements contribute to the total error by less than 5 \%.
As the situation might change throughout a time-domain simulation,
we consider the variable fraction marking the most suitable for our problems.

Also for the element marking distinct strategies are applied for constructing the initial
$hp$-mesh and during the transient analysis.
For generating the initial $hp$-mesh, we perform D\"{o}rfler-marking.
The number of mesh adaptation iterations required for obtaining the initial mesh depends on the
fraction of the total error. Less iterations are performed for large fractions. However, this
usually leads to a slightly larger number of DoF.

During the transient analysis a slightly altered marking strategy is employed.
This strategy is a variable fraction strategy with respect to the number of elements as well as
to the total error. For every mesh adaptation 
a minimal subset $\mathcal{T}_h^+$ is assembled such that
\begin{equation}
  \label{eq:markingDynamicRefine}
  \sum_{\Ti \in \mathcal{T}_h^+} \varepsilon_i^2 \ge \min \Big( \sum_{\Ti \in \mathcal{T}_h} \theta^2_+\varepsilon_i^2, \max\:(\varepsilon_{hp}^2 - \mathtt{TOL}^2,0) \Big).
\end{equation}
Hence, the size of the minimal subset of elements to
be refined is not larger than determined by the given fraction $\theta$, but it can be smaller
if the global error is close to the prescribed tolerance. If the estimated global error
is smaller than $\mathtt{TOL}$, the set is empty, and no elements are refined in this adaptation step.
If we were to apply the marking strategy in the same way we did for obtaining the initial
mesh, the algorithm would continue refining elements even if the estimated error
is less than the tolerance.

As stated above, we assume that the approximation on the current $hp$-mesh
is sufficiently accurate for serving as the reference solution. This statement should ideally
be true for every element. Therefore, marking elements for derefinement has to be
done with care.
We recall that mesh adaptation during the transient analysis is a dynamic process.
Therefore, elements suitable for derefinement, which are not marked as such
in an adaptation step are again considered for derefinement in the next step. In the examples
in Sec.~\ref{sec:examples}, we show that the mesh derefinement works
well despite the careful approach.

\subsection{The $hp$-Decision}
\label{sec:hp-decision}

Following the decision on which elements to adapt, the kind of adaptation
has to be chosen, i.e., $h$- or $p$-adaptation. This decision is guided
by the local solution smoothness. It is well known that for sufficiently smooth
solutions consecutive $p$-enrichment leads to exponential convergence, whereas
$h$-refinement yields algebraic convergence rates
only~\cite{Babuska:1994wj,Schwab:1999tu}. 

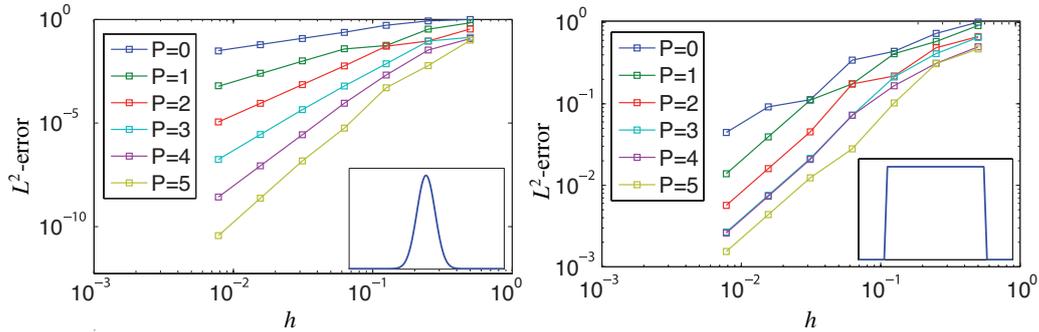
\begin{figure}[tb]
  \centering
  \includegraphics[width=0.8\linewidth]{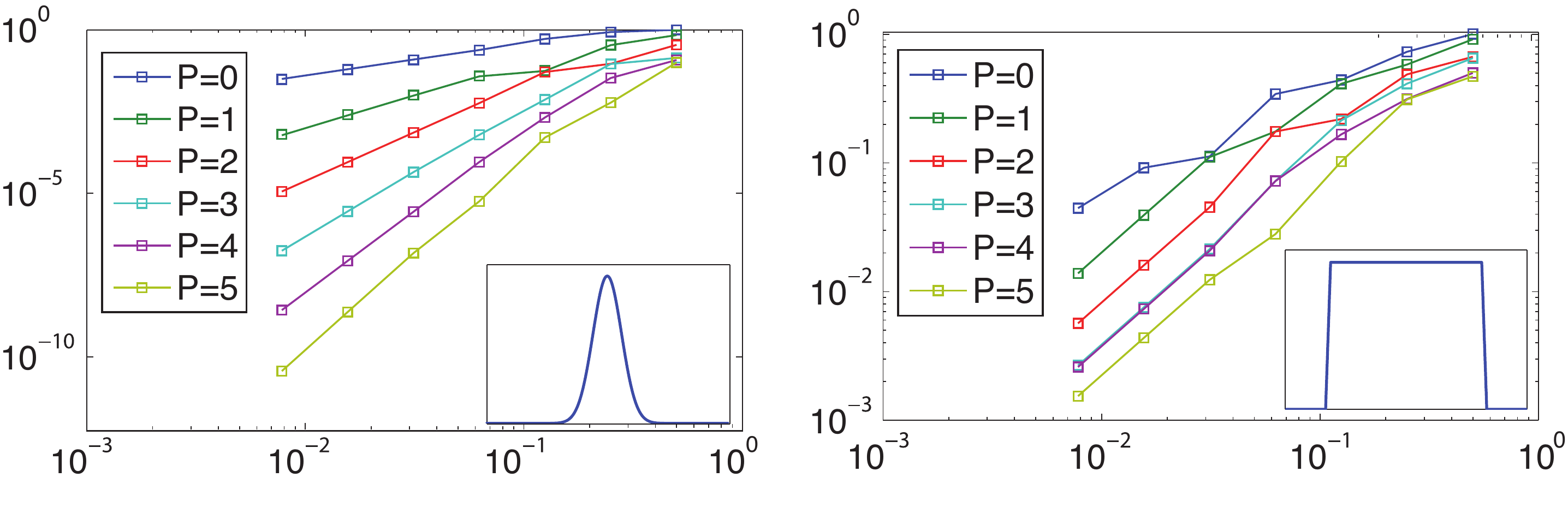}
  \begin{picture}(0,0)(0,0)
    \put(-388,52){\text{\small \begin{sideways}$L^2$-error\end{sideways}}}
    \put(-193,50){\text{\small \begin{sideways}$L^2$-error\end{sideways}}}
    \put(-90,2){\text{\small $h$}}
    \put(-285,2){\text{\small $h$}}
  \end{picture}
  \caption{Global projection error of a Gaussian and a trapezoidal waveform as depicted in the insets.
The plot in the left panel shows convergence of the error at a rate of $P+1$ for the Gaussian waveform.
In the right panel, the low regularity limits convergence to first order.}
  \label{fig:GaussRect}
\end{figure}

Figure~\ref{fig:GaussRect} illustrates the dependence of
the convergence rate on the regularity.
The waveforms depicted in the insets, i.e., a Gaussian and a
trapezoidal waveform in one-dimensional space, are projected
to spaces $\mathcal{V}^P$ with $P$ varying from zero to five.
The plots show the global error measured in the $L^2$-norm. While
the convergence rate increases from one to six with every increase of $P$ for the Gaussian waveform,
convergence is limited to first order in the latter case. 

However, exponential convergence in terms of DoF can be obtained even for
locally non-smooth solutions as well by employing proper
$hp$-refinement~\cite{Babuska:1994wj}. To this end, regions of low regularity
are embedded into $h$-refined areas of the mesh using low order
polynomials. Then, $p$-refinement is
applied everywhere else. Thus, the performance of the adaptive method
critically hinges on correct $hp$-decisions.
In order to be in the position of performing anisotropic $hp$-refinement
in three-dimensional space, we
require information about the directional smoothness of the unknown solution.

A variety of techniques have been proposed for assessing the
local smoothness or, more general, for making the $hp$-decision.
The simplest ones makes use of information
that is available \textit{a priori} such as the position of field singularities
due to edges and corners~\cite{Fahs2007}. However, instead of relying on geometric
information, we rather wish to drive the $hp$-decision based on the actual
numerical solution. Known methods include the type parameter technique~\cite{Gui:1986gz},
'Texas 3-Step'~\cite{Oden:1992tg}, mesh optimization techniques~\cite{Rachowicz:1989id},
error prediction~\cite{Melenk:2001cy} or local regularity
estimations~\cite{Ainsworth:1998gh,Houston:2003wv,Houston:2005p1984,Wihler:2011jf}.
Descriptions of all methods are beyond the scope of this
paper, and we refer to~\cite{Mitchell:2011vy} for an extended overview
including descriptions.

A particularly popular method is the estimation of the Sobolev regularity index $s$ in a local manner.
One such technique is described briefly in the following,
as it is illustrative for understanding why we pursue a different strategy.
We focus on~\cite{Houston:2003wv,Houston:2005p1984}, where the authors develop
such a strategy based on monitoring the decay rate of the sequence of
coefficients in the Legendre series expansion of the numerical solution.
The drawback of this method
and similar ones is that a certain number of coefficients is required for the
computation of the coefficient decay rate to be robust. Taking into account
that the Legendre coefficient of order zero provides information about 
the average in the element only, coefficients providing actual decay information
start with the order of one. Hence, decay rate estimations require
second order approximations as a minimum
although higher order approximations will make the method more robust.
Problems also occur if the solution exhibits a pronounced odd-even
characteristic~\cite{Klockner:2011ia} leading to an alternation
of small and large valued coefficients. The extension to problems in two- and
three-dimensional space is possible but not unique, and
the technique loses part of its clarity. As approximation orders
of at least two have to be applied in all directions, this leads to a significant
number of DoF also in elements, which do not require it.

As we wish to employ approximation orders as low as possible everywhere
the solution permits, we follow a different approach. 
To this end, we reuse the reference solution at hand for finding the
most suitable refinement from a list of candidates. This approach
circumvents the issue of regularity estimation and the associated difficulties
by testing various $h$-, $p$- and $hp$-candidates with respect
to the reference solution. With this strategy, the best candidate naturally arises
as the one offering the best ratio of approximation error $\varepsilon_c$ to
the logarithm of its number of DoF ($\varepsilon_c/\log(\#\text{DoF})$).

The size of the list of candidates can vary considerably. It depends
on the global refinement strategy, i.e. isotropic refinement only,
fully anisotropic, or anisotropic in one of $h$ and $p$ only, but
it also depends on the permissible increment and decrement
in the $h$-refinement level $\Delta L$ and $\Delta P$. In this
paper, we restrict both to one. However,
candidates have to be competitive. This means that increasing the
$h$-refinement level $L$, or $L_d$ in the anisotropic case, goes along
with a reduction of $P$ in order to prevent a strong increase of the
number of DoF in the element. The approximation order is reduced such that the number
of DoF of the candidate is as small as possible but larger than the
one of the current element ($\#\text{DoF}_c >  \#\text{DoF}_i$). If isotropic refinement is applied
two refinement candidates are obtained, one
$h$-candidate, consisting of eight elements with possibly
decreased approximation order $P$, and one $p$-candidate.
For fully anisotropic refinement a number of fourteen candidates
is considered, which are obtained by refining each of 
the mesh parameters $h$ or $p$ in one direction
(three candidates), two directions (three candidates), and all three
directions (one candidate).
The approximation order $P$ of $h$-candidates is reduced
as described above.
The procedure above applies to mesh refinement. For the case of
mesh derefinement, it is natural to proceed in a similar manner
and set up derefinement candidates with a smaller number of DoF.

In the dynamic case, the procedure requires modification
as the problem is encountered that a refined reference solution
cannot be constructed.
An error estimate is obtained by projecting to a $p$-reduced
FES, however, given our description of regularity estimation
based on coefficient decay rates,
it is doubtful that regularity information can be extracted from a comparison of
two solutions of the order $P$ and $P-1$ in a robust way.
As there is not enough information available for making a
reasoned $hp$-decision, the respective element is refined
uniformly in $h$ and $p$.
During the next mesh adaptation one or more of the refined elements
might be derefined again according to the estimated error.
Hence, it is during mesh derefinement only that
anisotropic adaptation occurs.

\subsection{Mesh Adaptation}
\label{sec:mesh-adaptation}

At this point sufficient information is available for performing error driven $hp$-adaptation, which
can be anisotropic in both mesh parameters $h$ and $p$.
Upon mesh adaptation the numerical approximation given on the
current $hp$-mesh $\Th$ has to be transferred to the adapted mesh
$\Th^*$. The objective is to find the best representation
of $\vec{u}_{hp}$ on $\Th^*$ with respect to the $L^2$-norm. 
For all adaptations ($h$/$p$ refine/derefine) this is achieved by applying the orthogonal projection operator $\Pi$
introduced in~\eqref{eq:projectionOperator}. Due to the
compact support of the basis, an unconstrained projection can be carried out
in a strictly element-wise fashion. Additionally, the tensor
product property of the basis~\eqref{eq:DGtensorSpace} allows
for performing the projection along each dimension individually.
This reduces the three-dimensional quadrature of complexity order three
in the number of quadrature nodes
to a product of three one-dimensional quadratures of complexity
order one. For the details of the projection we refer to \cite{Schnepp:2012JCAM}, where
the issues of optimality, stability and efficiency are investigated in details. 

\subsection{Comments on Practical Issues}
\label{sec:comm-pract-issu}

During one time-domain simulation a very large number of adaptations
is performed. These have to be administered in a way, which allows
for an efficient traversing of all elements in each time step. Additionally,
parent-child information is required for simplifying mesh derefinement.
In this context, tree structures emerge as a suitable storage format. They allow for operating
on the current discretization by working on the tree leaves only but
contain the refinement history and parental relationships as well.

In the case of isotropic $h$-refinement, the tree is organized using
octree-structures, where each of the eight children is assigned to one
branch. In an octree-structure, every element and its associated node
is either a non-reducible element of the root tesselation $\mathcal{T}_0$
or one of eight children of a single parent element. The depth of a node in the
tree, i.e. the number of ancestor elements to the respective root element,
corresponds to the number of consecutive $h$-refinements. This has been
defined as the $h$-refinement level $L$ before.

\begin{figure}[tb]
  \centering
  \includegraphics[height=2.5in]{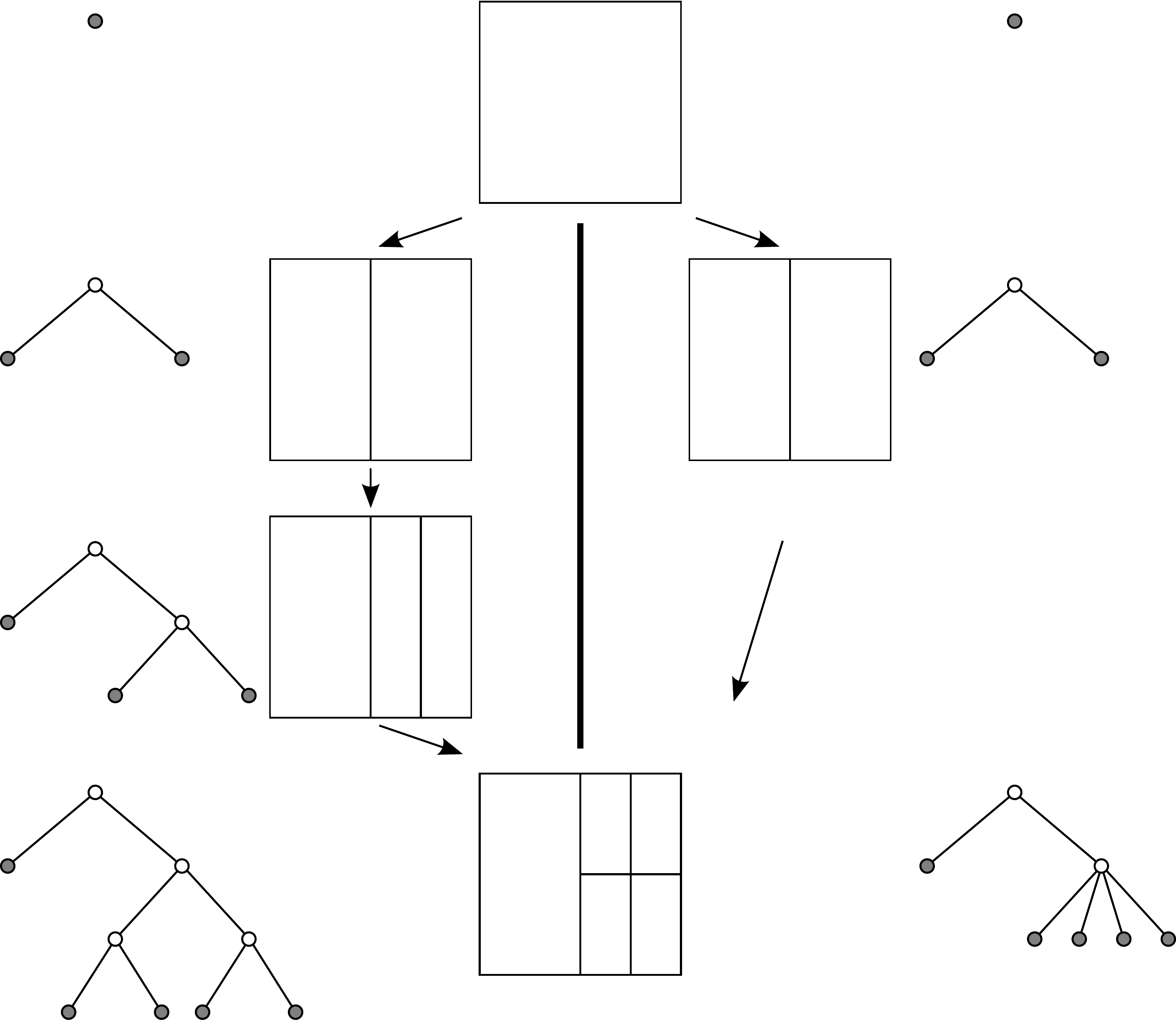}
    \put(-165,137){\text{\small aniso}}
    \put(-68,137){\text{\small aniso}}
    \put(-165,92){\text{\small aniso}}
    \put(-68,68){\text{\small iso}}
    \put(-165,46){\text{\small aniso}}
    \put(-201,183){\text{Tree I}}
    \put(-43,183){\text{Tree II}}
    \put(-117,108){\begin{turn}{270}\text{Option I}\end{turn}}
    \put(-102,73){\begin{turn}{90}\text{Option II}\end{turn}}
    \put(8,3){\line(0,1){175}}
    \hspace{6mm}
    \includegraphics[height=2.5in]{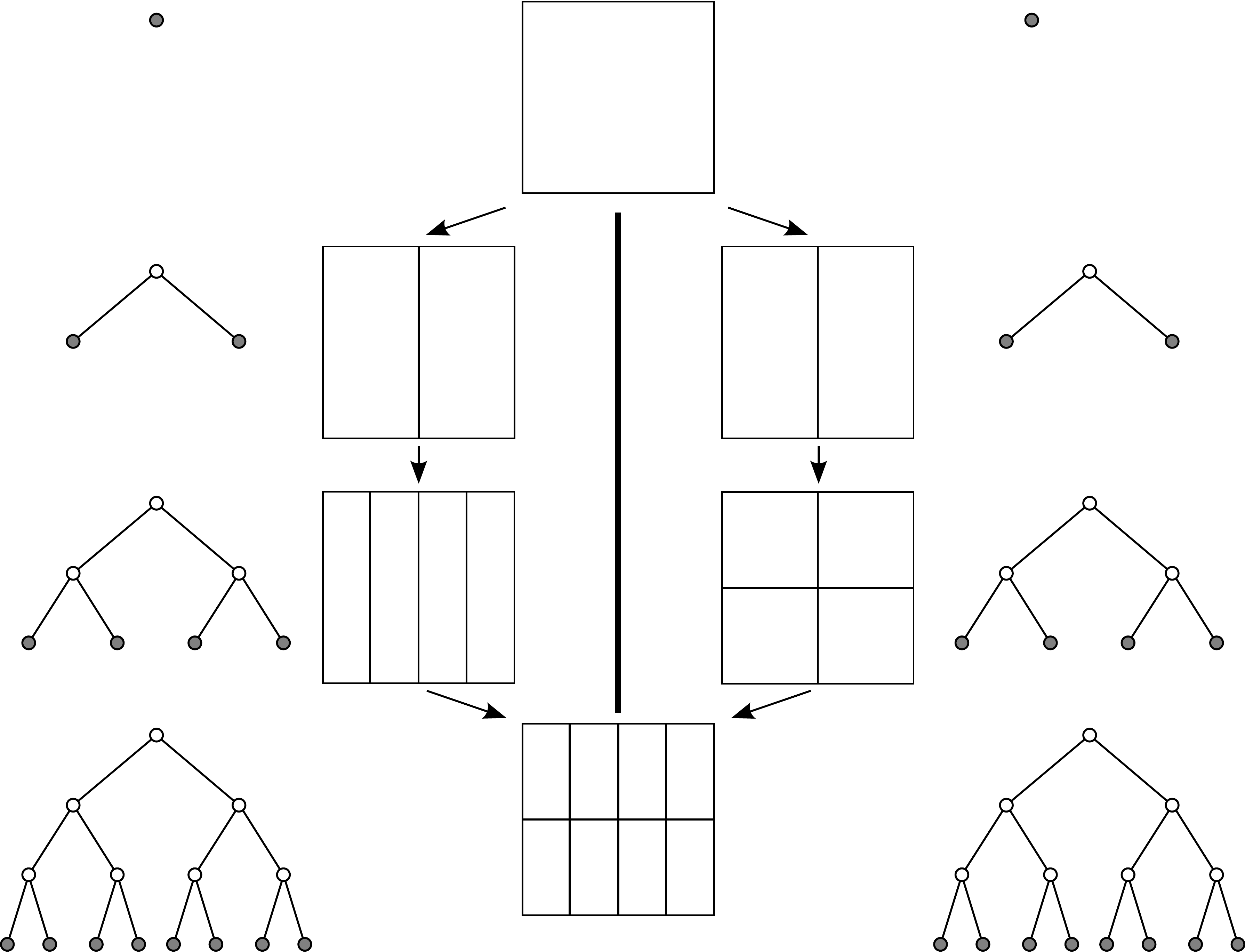}
    \put(-178,137){\text{\small aniso}}
    \put(-78,137){\text{\small aniso}}
    \put(-178,90){\text{\small aniso}}
    \put(-78,90){\text{\small aniso}}
    \put(-178,43){\text{\small aniso}}
    \put(-78,43){\text{\small aniso}}
    \put(-220,183){\text{Tree III}}
    \put(-58,183){\text{ Tree IV}}
    \put(-130,108){\begin{turn}{270}\text{Option I}\end{turn}}
    \put(-115,73){\begin{turn}{90}\text{Option II}\end{turn}}
  \caption{Comparison of mesh representation trees of different refinement histories.
    Starting from a single element (top) the final refinement at the bottom is obtained.
    For Option I in the left example only anisotropic refinement is applied, in Option II mixed
    anisotropic and isotropic refinement is employed.
    In this example, different representation trees are obtained for identical meshes.
    In the right hand example identical trees are obtained for identical final
    meshes although refinements were performed in a different order.
    Despite their identical appearance, trees III and IV differ, which becomes obvious when
    mesh derefinement is performed by cutting branches from the bottom up.
    These examples depict simple situations in two-dimensional space, 
    in three dimensions more options arise.
  }
  \label{fig:IsoVsAniso}
\end{figure}
\begin{figure}[tb]
  \centering
  \includegraphics[height=2.5in]{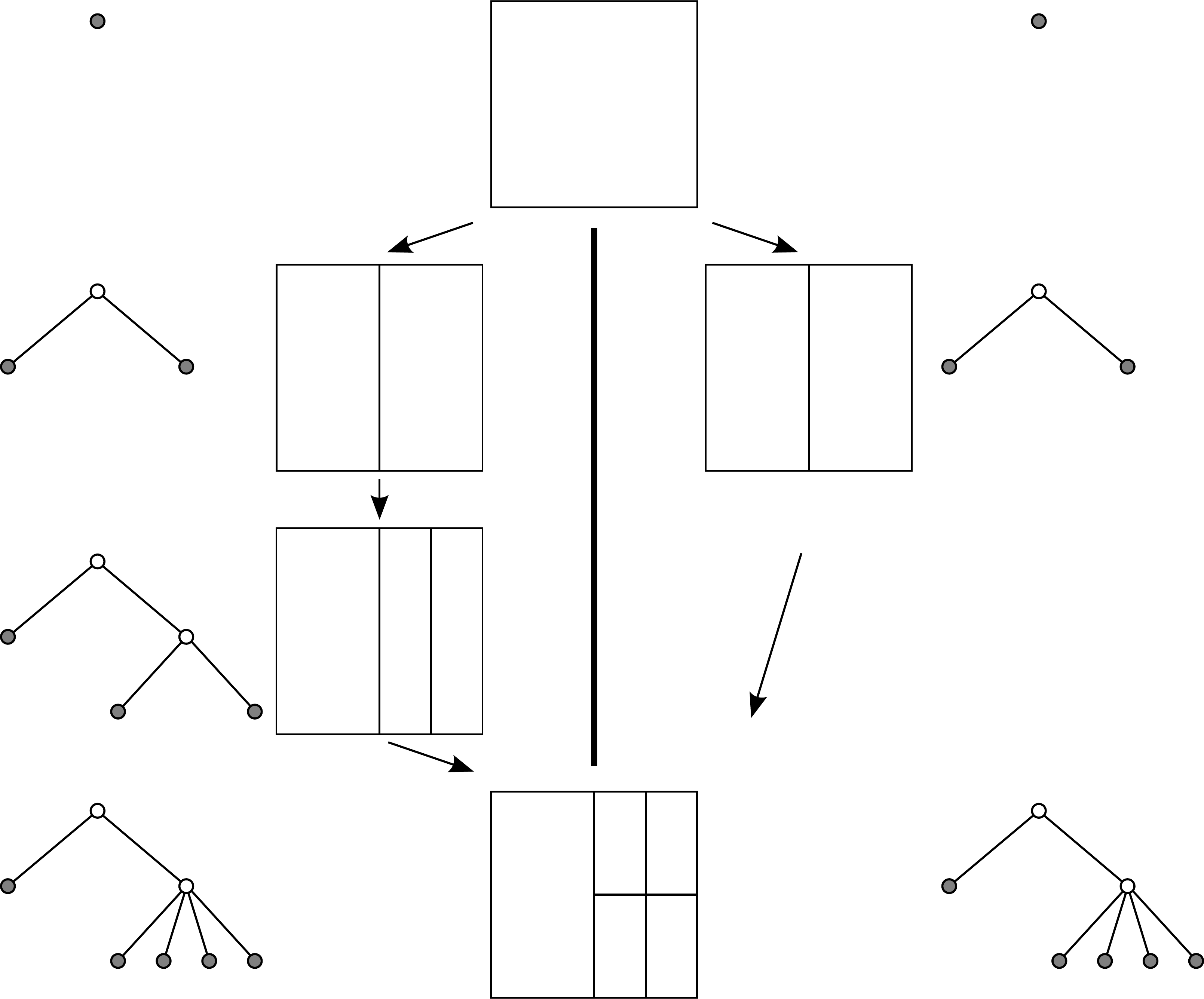}
    \put(-170,136){\text{\small aniso}}
    \put(-68,136){\text{\small aniso}}
    \put(-170,88){\text{\small aniso}}
    \put(-72,63){\text{\small iso}}
    \put(-170,41){\text{\small aniso}}
    \put(-221,183){\text{Min. Tree I}}
    \put(-53,183){\text{Min. Tree II}}
    \put(-122,108){\begin{turn}{270}\text{Option I}\end{turn}}
    \put(-107,73){\begin{turn}{90}\text{Option II}\end{turn}}
    \put(2,3){\line(0,1){175}}
    \hspace{1.5mm}
    \includegraphics[height=2.5in]{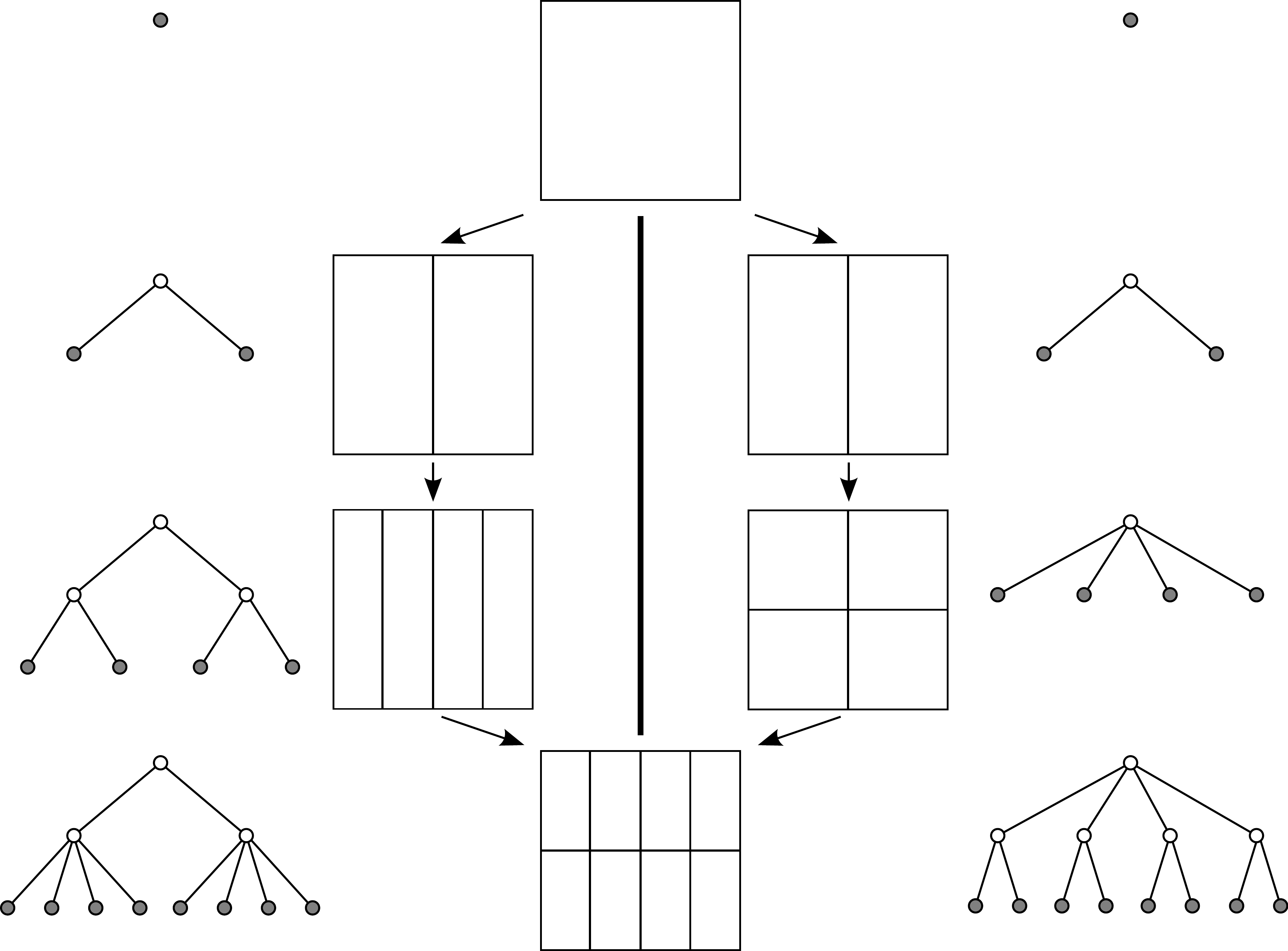}
    \put(-184,135){\text{\small aniso}}
    \put(-78,135){\text{\small aniso}}
    \put(-184,87){\text{\small aniso}}
    \put(-78,87){\text{\small aniso}}
    \put(-184,39){\text{\small aniso}}
    \put(-78,39){\text{\small aniso}}
    \put(-233,183){\text{Min. Tree III}}
    \put(-58,183){\text{ Min. Tree IV}}
    \put(-135,108){\begin{turn}{270}\text{Option I}\end{turn}}
    \put(-120,73){\begin{turn}{90}\text{Option II}\end{turn}}
  \caption{The same refinements as in Fig.~\ref{fig:IsoVsAniso} are performed. The
    representation tree is constructed following a minimal depth strategy. With this strategy
    the maximum depth of the representation tree corresponds to the maximum of the refinement levels $L_d$.
    Trees obtained with this strategy provide more freedom for performing mesh derefinement.
  }
  \label{fig:MinTree}
\end{figure}
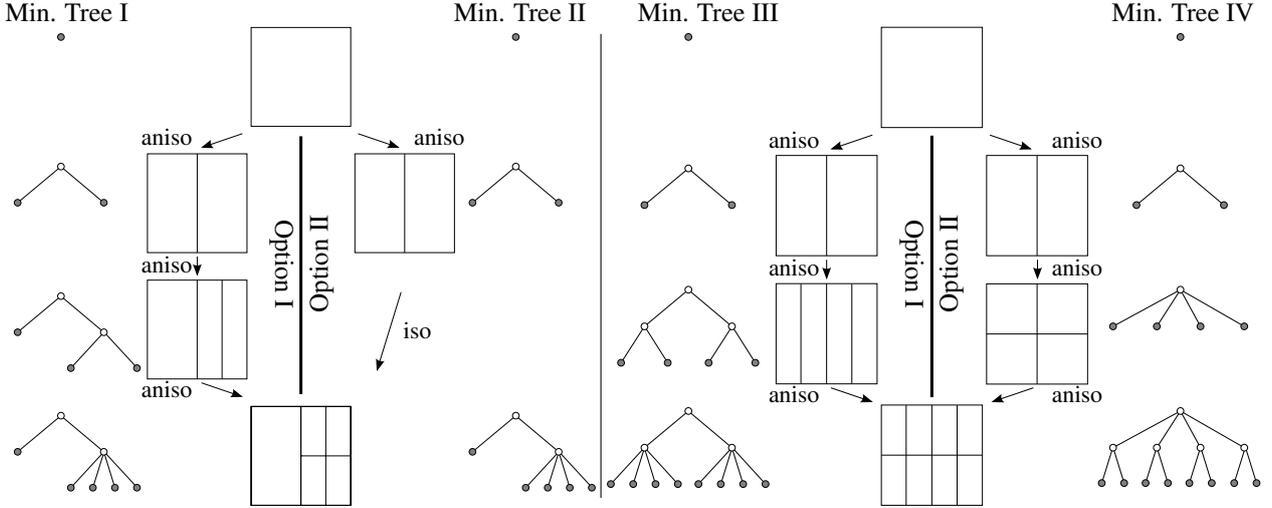

This organized view breaks down if non-anisotropic refinement is permitted.
We refer to Fig.~\ref{fig:IsoVsAniso} for the following explanation.
For the sake of clarity, a single two-dimensional element is considered.
In Option I of the left hand side example only anisotropic $h$-refinement is applied. 
We extend the mesh representation tree in the
same way as for isotropic refinement, i.e., the splitting of elements 
for every $h$-refinement is represented by extending the tree downwards from
the respective node. In Option II, a combination of 
anisotropic and isotropic $h$-refinement is performed. This yields
the same final mesh but a different representation tree.
For the example on the right of Fig.~\ref{fig:IsoVsAniso}, the same
refinements are performed in a different order leading to identical
final results and apparently identical representation trees.

The problem associated with the representation trees in Fig.~\ref{fig:IsoVsAniso}
becomes visible when we attempt to derefine the mesh. 
This is achieved by cutting branches from 
the leaves upwards to the root, which immediately implies that derefinement has to occur exactly in the
reversed refinement order. This is not
a desired behavior, as the solution can develop in a way such
that a different derefinement order would be more suitable.
We point out, that the simple two-dimensional examples
of Fig.~\ref{fig:IsoVsAniso} suggest that this is a minor
issue. Nevertheless, in more complex situations in three-dimensional
space it is a clear disadvantage if mesh refinement and derefinement
have to be performed in reversed order.

The issue can be faced in a number of ways, many of them
being computationally expensive. As one example, graph theory
could be applied for generating a new minimal representation tree
after each refinement. Our approach is computationally much cheaper
and aims at constructing representation trees of minimal depth.
The idea is illustrated in Fig.~\ref{fig:MinTree}. In order to obtain
a tree of minimal depth, a new generation of children is spawned
only if the maximum $h$-refinement level $L = \max (L_d)$ is increased.
For the minimal tree I (Min. Tree I), this is the case for the first two refinements
but not for the last refinement step. This strategy yields
identical minimal representation trees I and II. However, the uniqueness of
minimal trees is not guaranteed by the approach as demonstrated in the right 
hand example of Fig.~\ref{fig:MinTree}.
Nevertheless, for general refinements in three dimensions trees of a significantly
smaller depth are obtained. They also provide a more intuitive representation as
the tree depth connects with the maximum $h$-refinement level.

The important benefit of constructing minimal trees becomes evident when
mesh derefinement is considered. In contrast to the trees constructed
in Fig.~\ref{fig:IsoVsAniso}, the derefinement order is not strictly prescribed by
the refinement order. The minimal tree I allows for derefining such that
the meshes at steps one or two are obtained. Additionally, a mesh with one horizontal
and one vertical refinement of the right hand side element is obtained naturally.
Using minimal trees, identical representations, such as I and II, always offer identical
derefinement options, which is a significant advantage regarding the
implementation in a computer code.
Considering minimal tree III all meshes depicted in either of the options I or II
can be obtained by derefinement.

\begin{figure}[tb]
  \centering
  \includegraphics[width=2.5in]{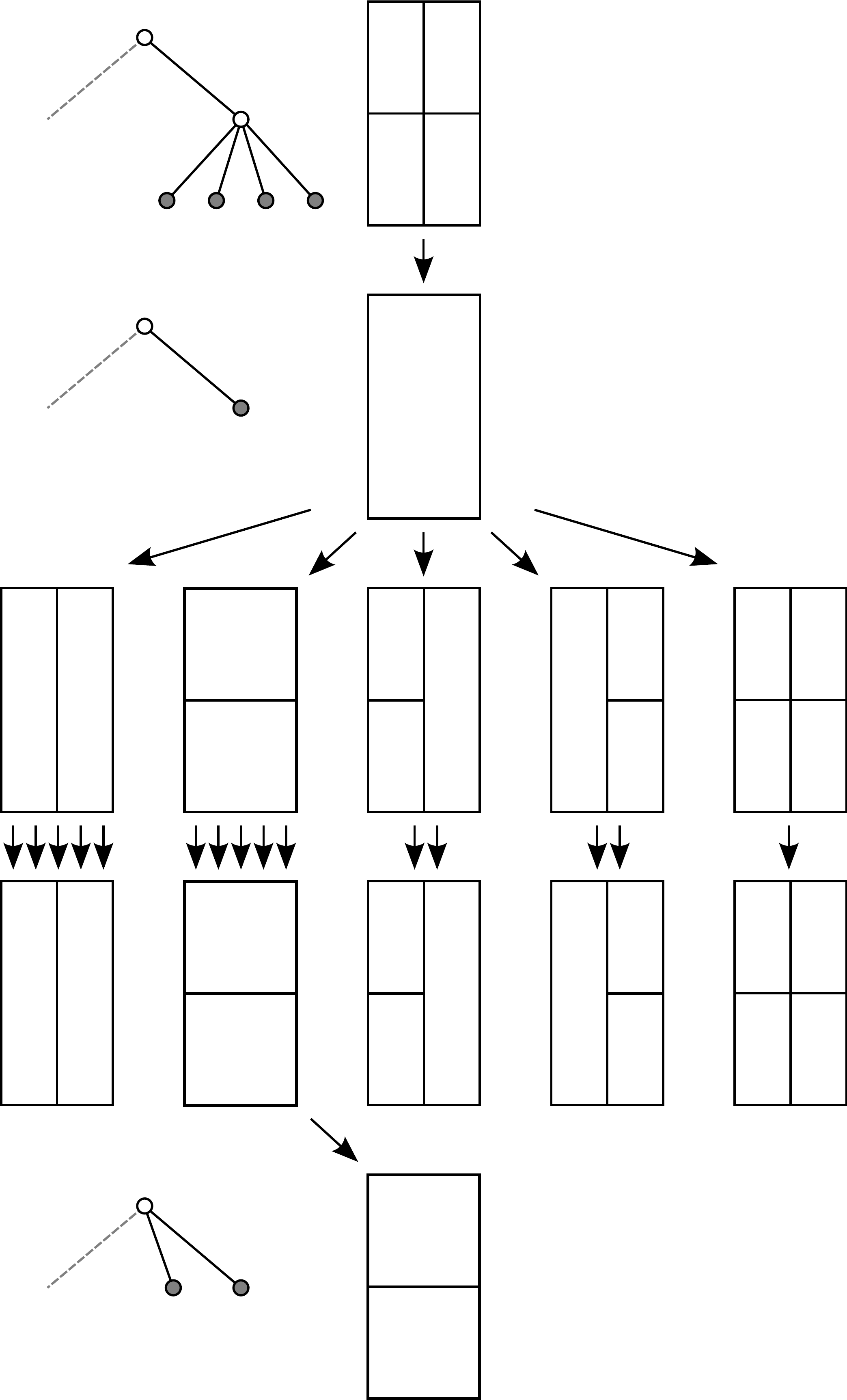}
    \put(-205,257){\begin{turn}{90}\text{Current}\end{turn}}
    \put(-195,260){\begin{turn}{90}\text{Mesh}\end{turn}}
    \put(-205,190){\begin{turn}{90}\text{Topological}\end{turn}}
    \put(-195,200){\begin{turn}{90}\text{Parent}\end{turn}}
    \put(-200,125){\begin{turn}{90}\text{$h$-Candidates}\end{turn}}
    \put(-200,55){\begin{turn}{90}\text{$hp$-Candidates}\end{turn}}
    \put(-200,13){\begin{turn}{90}\text{Winner}\end{turn}}
    \put(-95,8){\text{$P$}}
    \put(-95,32){\text{$P$}}
  \caption{Illustration of the derefinement strategy using the right hand side
    example of Fig.~\ref{fig:MinTree} and minimal tree III 
    as the starting point. Only the right hand half is depicted. In step one the topological parent is obtained. Next, the parent
    is $h$-refined again in order to generate a list of $h$-candidates. In step three $hp$-candidates are created by 
    using different approximation orders for each $h$-candidate. This step is restricted in the sense that all elements of one
    candidate have same order $P$. Also, each candidate has to have less DoF than the current mesh. All candidates
    are tested yielding the winning $hp$-derefinement option.
  }
  \label{fig:DerefinementAlgo}
\end{figure}
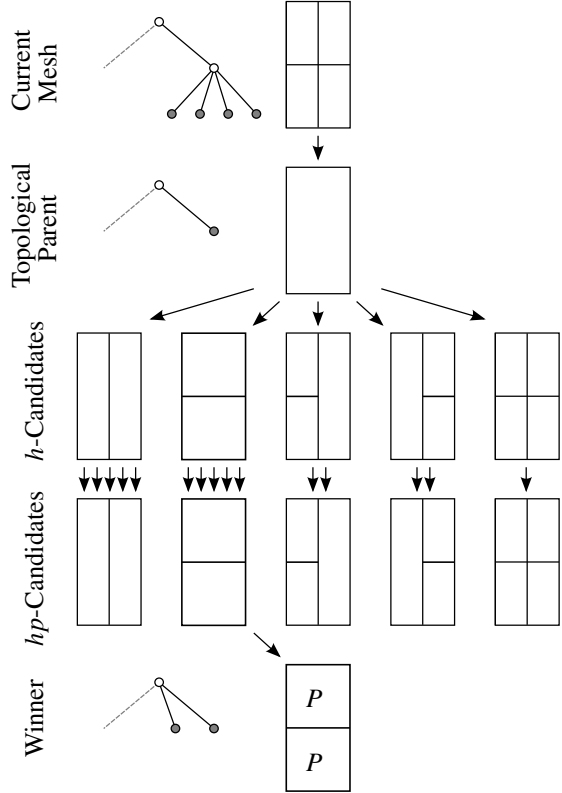

The selection algorithm for the most suitable derefinement candidate
is depicted in Fig.~\ref{fig:DerefinementAlgo}, where the mesh of
the right hand example in Fig.~\ref{fig:MinTree} and minimal tree III
is considered. Only the right hand half is depicted
as derefinement of the other half is carried out analogously.
Given the current discretization and its representation tree,
we move one level upwards in order to obtain the topological parent element.
The parent is the first $h$-derefinement candidate. Then, successively all
possible $h$-refinements of the parent are performed such that $L_{d,c} \le L_d$ is respected.
The additional $h$-candidates for the considered example are depicted
in the third row of Fig.~\ref{fig:DerefinementAlgo}. 
In a third step, the purely topological $h$-candidates are assigned
FES of different orders $P$ yielding $hp$-candidates.
We restrict the generation of $hp$-candidates in the sense that all
elements of a candidate have the same order $P$.
However, each $h$-candidate has its own $P$ dictated
by the requirement that the number of DoF of the candidate has to be smaller
than in the current mesh. In a last step, we compute $\varepsilon_c/\log(\#\text{DoF})$
for each $hp$-candidate and choose the best derefinement option.

\section{Examples}
\label{sec:examples}

\subsection{Propagation in a waveguide}
\label{sec:fund-mode-waveg}

As a first example we consider the propagation of a wave packet
in a rectangular waveguide. We consider a waveguide of type WR 19 working in U-Band.
The cutoff frequency of the fundamental mode is 40 GHz. The frequency limit for single-mode operation
is 60 GHz, and the wave packet considered has a frequency range of 45-59 GHz.
The waveguide aperture dimensions are $4.78 \times 2.39$~mm, and we consider
a total length of 1~m corresponding to approximately 170 wavelengths. 
The purpose of this rather academic example is to demonstrate that
the proposed algorithm can cope with situations where a very large
number of adaptations has to be performed. Throughout the simulation
the error tolerance has to be respected. Also, the number of DoF should
remain approximately constant as the wave packet will largely keep its shape.

\begin{figure}[tb]
  \centering
  \includegraphics[width=5.5in]{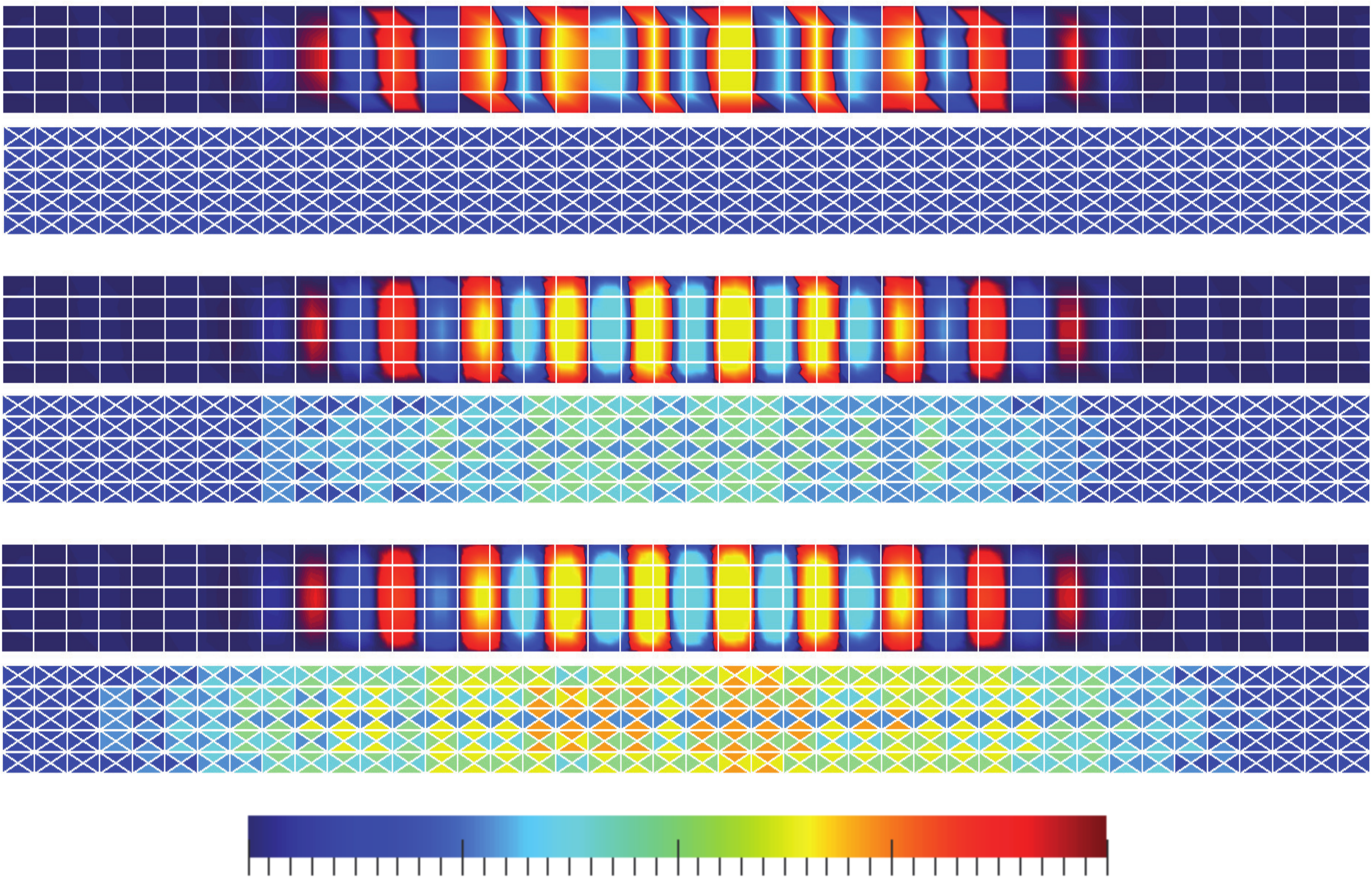}
    \put(-418,197){\begin{turn}{90}\text{Iteration \#0}\end{turn}}
    \put(-418,115){\begin{turn}{90}\text{Iteration \#14}\end{turn}}
    \put(-418,36){\begin{turn}{90}\text{Iteration \#28}\end{turn}}
    \put(-407,76){\begin{turn}{90}\text{\footnotesize $E_y$}\end{turn}}
    \put(-407,30){\begin{turn}{90}\text{\footnotesize $hp$-mesh}\end{turn}}
    \put(-407,155){\begin{turn}{90}\text{\footnotesize $E_y$}\end{turn}}
    \put(-407,110){\begin{turn}{90}\text{\footnotesize $hp$-mesh}\end{turn}}
    \put(-407,233){\begin{turn}{90}\text{\footnotesize $E_y$}\end{turn}}
    \put(-407,188){\begin{turn}{90}\text{\footnotesize $hp$-mesh}\end{turn}}
    \put(-415,103){\line(1, 0){422}}
    \put(-415,181){\line(1, 0){422}}
    \put(-240,-18){\text{\small polynomial order $p$}}
    \put(-327,-8){\text{\footnotesize  0}}
    \put(-264.7,-8){\text{\footnotesize  2}}
    \put(-202.5,-8){\text{\footnotesize  4}}
    \put(-140.3,-8){\text{\footnotesize  6}}
    \put(-78,-8){\text{\footnotesize  8}}
    \put(-401,-5){\vector(1, 0){15}}
    \put(-401,-5){\vector(0, 1){15}}
    \put(-400,11){\text{\footnotesize  $x$}}
    \put(-388,-2){\text{\footnotesize  $z$}}
  \caption{Generation of the initial $hp$-mesh for a Gauss-modulated sinusoidal waveform in the fundamental mode of a rectangular
    waveguide using anisotropic refinement. The $y$-component of the electric field
    and the $hp$-mesh is depicted in a cut view of a short waveguide section. The mesh is adapted iteratively such that
    the approximation error respects the tolerance $\mathtt{TOL} = 10^{-5}$ in the global $L^2$-norm. 
    Iteration \#0 shows the approximation on the root tesselation and the initially uniform polynomial order. 
    The adaptation terminates after 28 iterations, obtaining the initial $hp$-mesh depicted at the bottom. 
    The autonomous adaptation algorithm employs $p$-enrichment only, which is desirable as the solution is smooth.
    The $hp$-meshes are depicted using a common tensor product visualization technique based on embedded triangles.
    The highest order $P_z$ employed in the initial $hp$-mesh is
    six. The respective $z$-oriented edges are part of orange colored triangles. The maximum of $P_x$ is
    five (yellow).
  }
  \label{fig:wgInit}
\end{figure}

The generation of the initial $hp$-mesh required 28 iterations with the fraction $\theta$ as
described in~\eqref{eq:markingDynamicRefine} set to 0.5. The series depicted in Fig.~\ref{fig:wgInit} 
shows the $hp$-mesh and the respective approximation of the $E_y$-component on
the uniform root tesselation, at an intermediate iteration and the final $hp$-mesh.
Refinement is allowed to be anisotropic in both mesh parameters, $h$ and $p$, though
the algorithm applies no $h$-refinement in this case. This is reasonable as the solution is
smooth. For depicting anisotropic $hp$-meshes we make use of a common visualization
technique \cite{Demkowicz1,Solin2008117}. To this end, each face is split into
four triangles. The tensor product orders are coded with the triangle color. If
the base edge of a green triangle is aligned with the $x$-axis, then $P_x = 4$
according to the color legend. In the same way $P_z=5$ is represented with an orange triangle
having its base edge aligned with the $z$-axis. This visualization
allows for representing the orders in one plot and also gives an immediate 
impression of predominant directions regarding the approximation orders.

\begin{figure}[bt]
  \centering
  \vspace{-3mm}
  \includegraphics[width=3.5in]{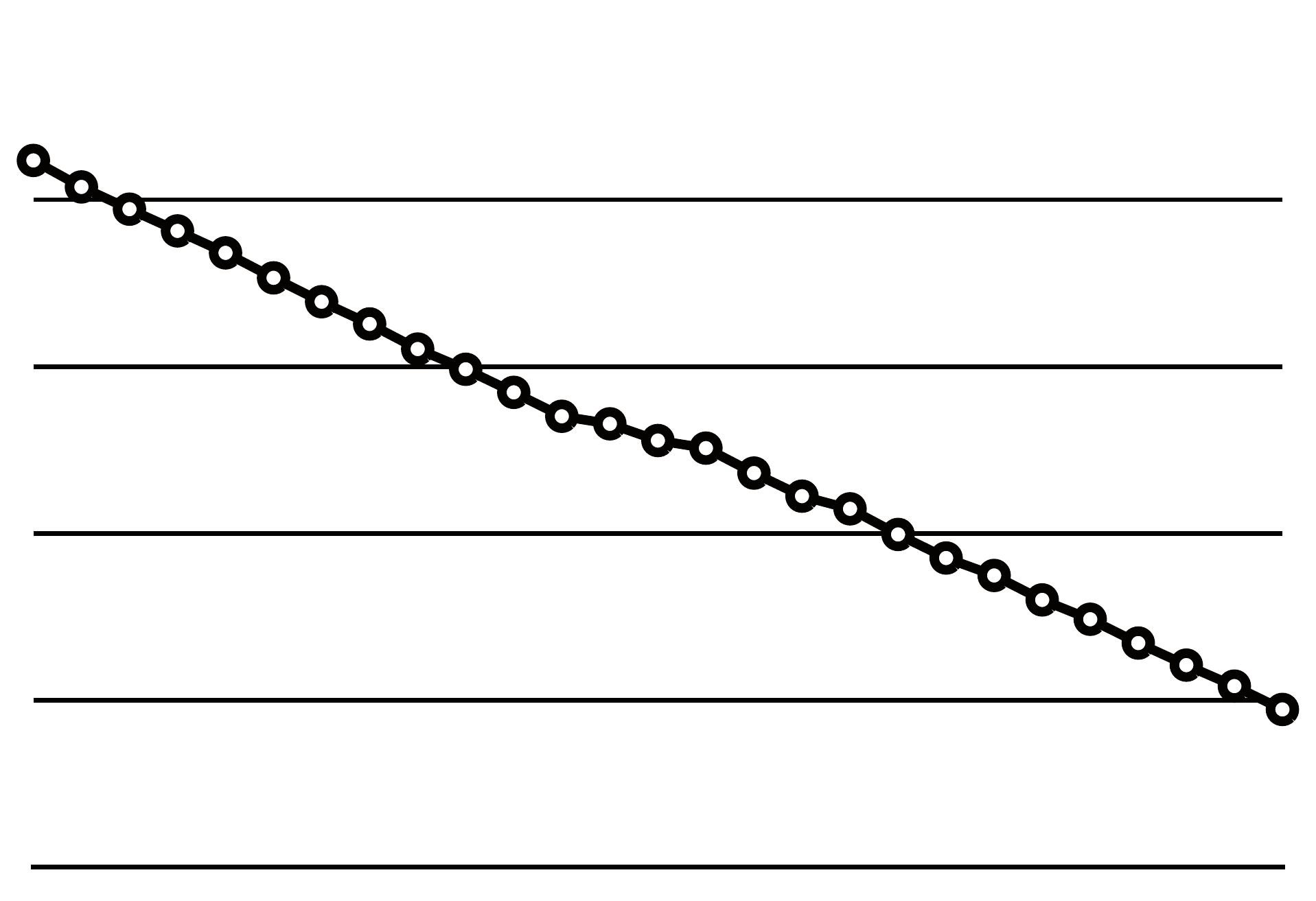}
    \put(-135,-18){\text{\small \#DoF / $10^3$}}
    \put(-245,-4){\text{\small  35}}
    \put(-199,-4){\text{\small  40}}
    \put(-152,-4){\text{\small  45}}
    \put(-106,-4){\text{\small  50}}
    \put(-60,-4){\text{\small  55}}
    \put(-12,-4){\text{\small  60}}
    \put(-280,45){\begin{turn}{90}\text{\small norm.~$L^2$-error}\end{turn}}
    \put(-265,133){\text{\small  $10^{-2}$}}
    \put(-265,101){\text{\small  $10^{-3}$}}
    \put(-265,69){\text{\small  $10^{-4}$}}
    \put(-265,36){\text{\small  $10^{-5}$}}
    \put(-240,5){\line(0, 1){3}}
    \put(-194,5){\line(0, 1){3}}
    \put(-147,5){\line(0, 1){3}}
    \put(-101,5){\line(0, 1){3}}
    \put(-55,5){\line(0, 1){3}}
    \put(-7,5){\line(0, 1){3}}
    \thicklines
    \put(-247,6){\vector(1, 0){255}}
  \caption{Convergence of the global $L^2$-error
    during construction of the initial $hp$-mesh as depicted in
    Fig.~\ref{fig:wgInit}. The graph uses a logarithmic scale for the error and a linear one
    in the number of DoF, thus showing exponential convergence. The error tolerance of $10^{-5}$ in the global $L^2$-norm
    is met after 28 iterations.
  }
  \label{fig:ConvInitial}
\end{figure}

The highest orders in the initial mesh are $P_x = 5$ and $P_z = 6$. As the fundamental mode
shows no variation in $y$-direction, no increase of $P_y$ occurs. The construction of
the initial mesh requires seven seconds and yields close to 135,000 DoF. If we allow for isotropic refinement
only, an initial mesh with 285,600 DoF is obtained within twelve seconds.
Figure~\ref{fig:ConvInitial} shows the convergence graph of the approximation error
with the number of DoF in a semi-logarithmic plot.
In this graph, the error reduction occurs along an almost straight line
showing exponential convergence.

\begin{figure*}[tb]
  \centering
  \includegraphics[width=5.5in]{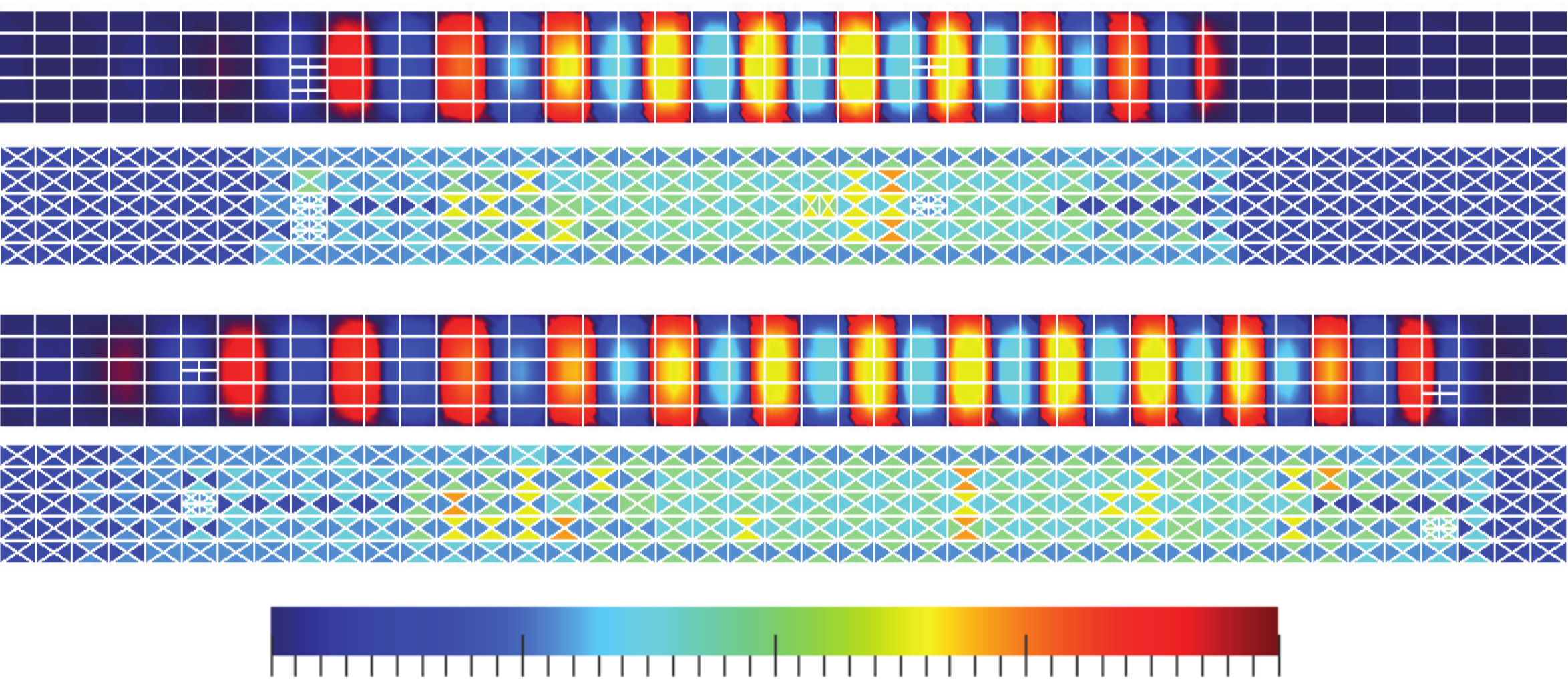}
    \put(-418,115){\begin{turn}{90}\text{$z = 0.5$ m}\end{turn}}
    \put(-418,37){\begin{turn}{90}\text{$z = 1.0$ m}\end{turn}}
    \put(-407,74){\begin{turn}{90}\text{\footnotesize $E_y$}\end{turn}}
    \put(-407,30){\begin{turn}{90}\text{\footnotesize $hp$-mesh}\end{turn}}
    \put(-407,151){\begin{turn}{90}\text{\footnotesize $E_y$}\end{turn}}
    \put(-407,107){\begin{turn}{90}\text{\footnotesize $hp$-mesh}\end{turn}}
    \put(-415,99){\line(1, 0){422}}
    \put(-240,-18){\text{\small polynomial order $p$}}
    \put(-330,-8){\text{\footnotesize  0}}
    \put(-266,-8){\text{\footnotesize  2}}
    \put(-202.5,-8){\text{\footnotesize  4}}
    \put(-139,-8){\text{\footnotesize  6}}
    \put(-75,-8){\text{\footnotesize  8}}
    \put(-401,-5){\vector(1, 0){15}}
    \put(-401,-5){\vector(0, 1){15}}
    \put(-400,11){\text{\footnotesize  $x$}}
    \put(-388,-2){\text{\footnotesize  $z$}}
  \caption{Evolution of the dynamic $hp$-mesh. Mesh refinement occurs predominantly in the form
    $p$-refinement. The snapshot show the field and mesh at the middle and end of the waveguide.
  }
  \label{fig:wgDyna}
\end{figure*}

\begin{figure}[tb]
  \centering
  \includegraphics[width=3.5in]{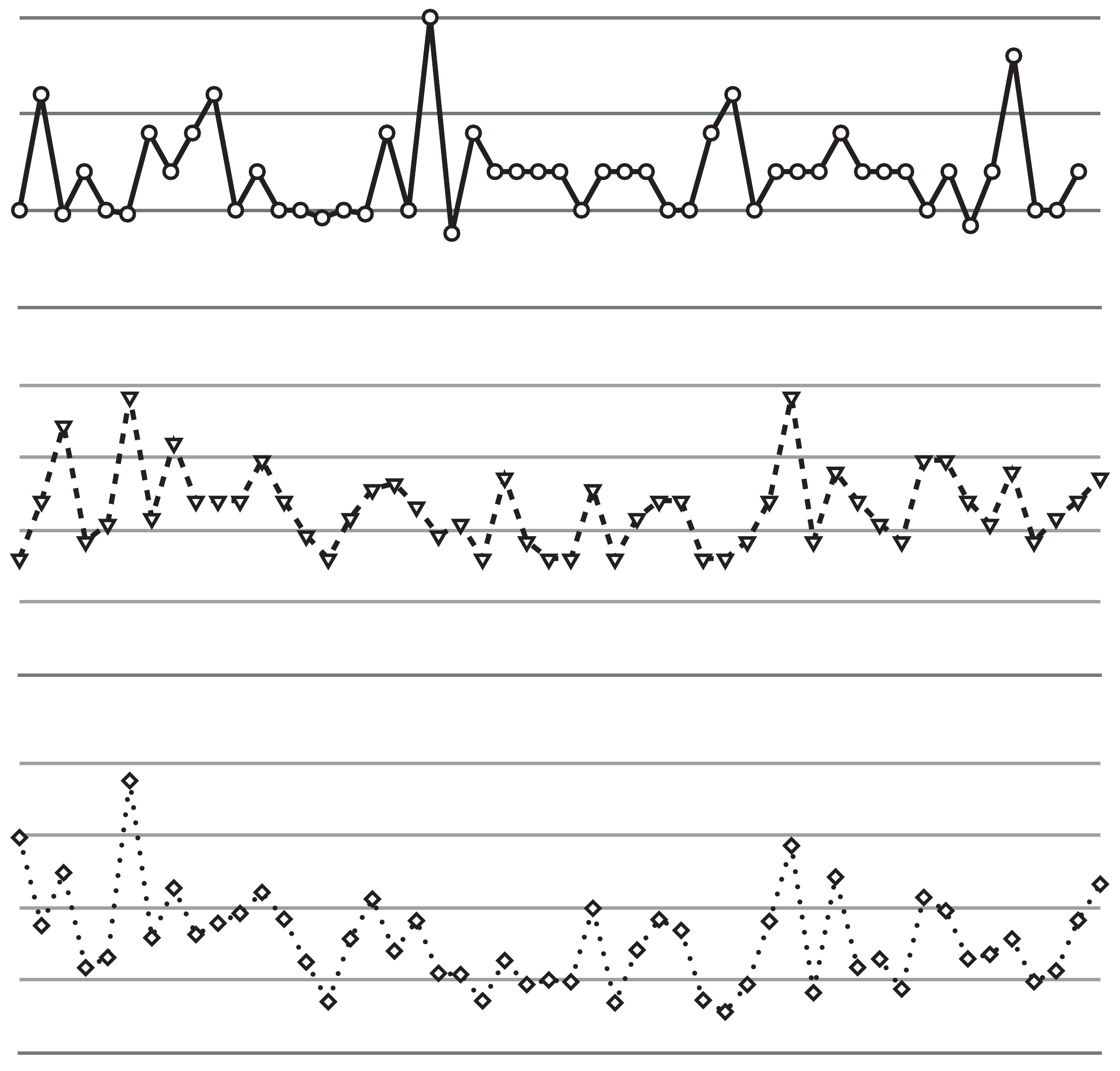}
    \put(-285,188){\begin{turn}{90}\text{\small norm.~error}\end{turn}}
    \put(-285,105){\begin{turn}{90}\text{\small \#elements}\end{turn}}
    \put(-285,20){\begin{turn}{90}\text{\small \#DoF / $10^3$}\end{turn}}
    \put(-270,20){\text{\footnotesize 125}}
    \put(-270,35.5){\text{\footnotesize 130}}
    \put(-270,51){\text{\footnotesize 135}}
    \put(-270,66.5){\text{\footnotesize 140}}
    \put(-268,103){\text{\footnotesize 2200}}
    \put(-268,119){\text{\footnotesize 2225}}
    \put(-268,135){\text{\footnotesize 2250}}
    \put(-268,151){\text{\footnotesize 2275}}
    \put(-270,192){\text{\footnotesize 1.000}}
    \put(-270,208){\text{\footnotesize 1.025}}
    \put(-270,224){\text{\footnotesize 1.050}}
    \thicklines
    \put(-249,89.5){\vector(1, 0){260}}
    \put(-249,172.5){\vector(1, 0){260}}
    \put(-139,-8){\text{time}}
    \put(-249,4){\vector(1, 0){260}}
  \caption{Temporal profiles of the global error normalized to the initial error (top, solid line with circles), 
    number of elements (middle, dashed with triangles) and number of degrees of freedom (bottom, dotted line with diamonds).
    The time range covers the full time-domain simulation sampled at 50 instances.
  }
  \label{fig:hpVStime}
\end{figure}

Next, the time-domain simulation is performed. Figure~\ref{fig:wgDyna} shows the
$E_y$-component and the $hp$-mesh after the packet has traveled to the
center and to the end of the waveguide.
The performance of the adaptive algorithm is illustrated in
Fig.~\ref{fig:hpVStime}. The top plot shows the evolution of the estimated global
$L^2$-error normalized to the error obtained on the initial mesh. The middle and bottom
plot depict the number of elements and DoF throughout the simulation. The data
corresponds to 50 samples in time. The dispersion, which can be observed in
Fig.~\ref{fig:wgDyna} is a physical effect due to waveguide dispersion, not a numerical artifact.
Code profiling showed that about 15~\% of the computing time is spent for
adaptation related tasks, which is almost negligible.

In order to assess the reduction in computing time and memory consumption due to adaptivity,
simulations on fixed meshes were carried out. The number of DoF, memory consumption, 
runtime and error estimates after the final time step for various settings
are listed in Tab.~\ref{tab:hpVSfixed}. In comparison to the adaptive solution
factors of about three to six are observed regaring computing time and memory
consumption on meshes using anisotropic approximation orders. For isotropic
orders these factors increase to about 20.


\begin{table}[bt]
  \renewcommand{\arraystretch}{1.3}
  \label{tab:hpVSfixed}
  \centering
  \begin{tabular}{r p{1.5cm} r r r r}
    \hline\hline
    \# & Orders \par ($P_x/P_y/P_z$)  & DoF / $10^3$ &  Memory / MB  &  norm. Runtime & $L^2$-error / $10^{-5}$\\
    \hline
     1 & \multicolumn{1}{r}{5/1/6} & {1131} & {35.5} & {4.3} & {0.13}  \\
     2 & \multicolumn{1}{r}{4/1/5} & {808} & {30.4} & {2.7} & {1.36}  \\
     4 & \multicolumn{1}{r}{5/5/5} & {2911} & {62.7}  & {10.3} & {1.31}  \\
     5 & \multicolumn{1}{r}{6/6/6} & {4620} & {88.7}  & {20.6} & {0.13}  \\
     6 & \multicolumn{1}{r}{$hp$}  & {125-140} & {4.7-5.3} & {1} & {1.01}\\
    \hline\hline
  \end{tabular}
  \caption{Performance of simulations of example~\ref{sec:fund-mode-waveg} using fixed and adaptive meshes. 
    The orders 5/1/6 correspond to highest orders obtained in the initial mesh (cf.~Fig.~\ref{fig:wgInit}). If these
    orders are used globally the error is much smaller than that of the adaptive solution. Reducing
    the orders to 4/1/5, however, exceeds the error of the adaptive solution by about 30 \%. Memory consumption
    of these fixed mesh solutions exceeds the adaptive solution by factors of 7 and 6, runtime by
    factors of 4 and 3. This has to be put in relation with the fraction of the mesh that is being refined,
    which corresponds to approximately 8 \% of the waveguide length.
    For comparison, two simulations using
    uniform orders of five and six are included, which leads to a significant increase of the number
    of DoF and runtime.
}
\end{table}

\subsection{Folded patch antenna}
\label{sec:folded-patch-antenna}

\begin{figure}[tb]
  \centering
  \includegraphics[width=3.5in]{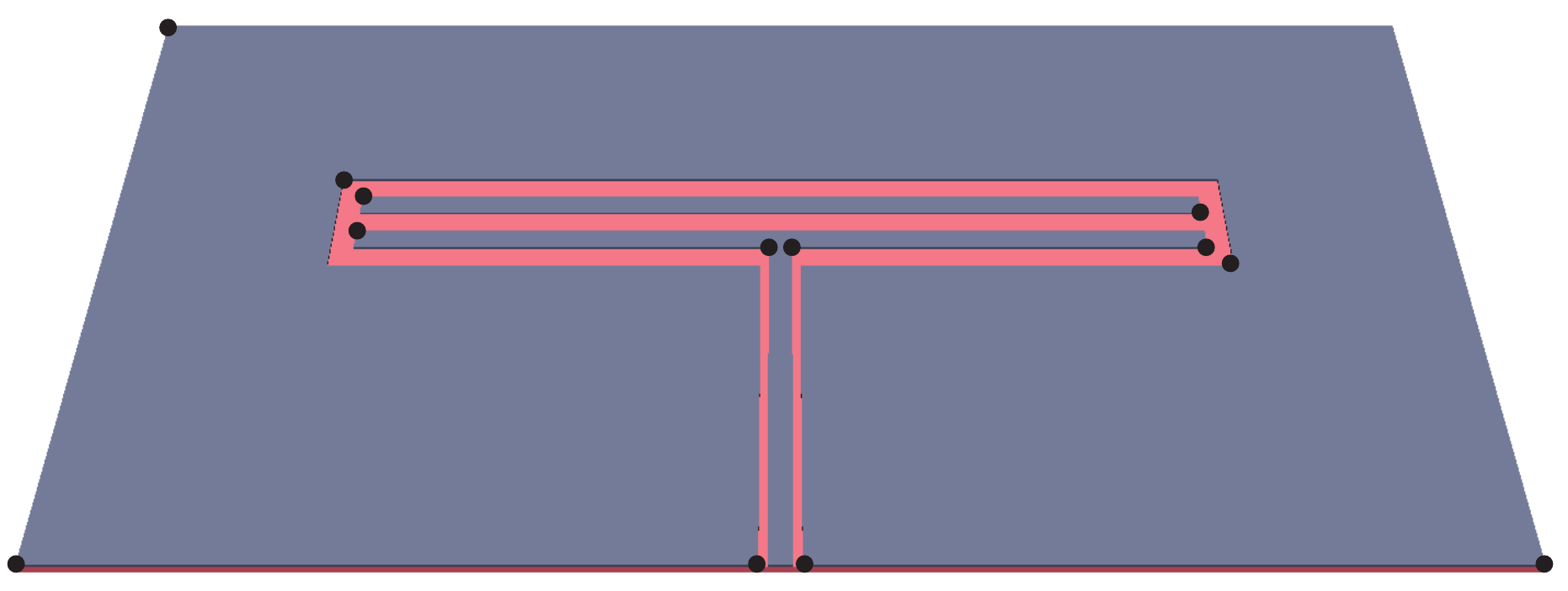}
    \put(-265,5){\text{\footnotesize 1}}
    \put(-138,5){\text{\footnotesize 2}}
    \put(-119,5){\text{\footnotesize 3}}
    \put(0,5){\text{\footnotesize 4}}
    \put(-50,50){\text{\footnotesize 5}}
    \put(-137,48){\text{\footnotesize 6}}
    \put(-119,48){\text{\footnotesize 7}}
    \put(-65,53){\text{\footnotesize 8}}
    \put(-190,53){\text{\footnotesize 9}}
    \put(-69,59){\text{\footnotesize 10}}
    \put(-193,61){\text{\footnotesize 11}}
    \put(-208,65){\text{\footnotesize 12}}
    \put(-242,87){\text{\footnotesize 13}}
  \caption{Triple slot folded patch antenna fixed on a dielectric substrate. 
    The positions of the points in the $x$--$y$-plane are given in Tab.~\ref{tab:patchsetup}.
    The relative permittivity of the substrate is 2.2. The thicknesses of the substrate and
    the metallization are 0.813 mm and 0.2 mm, respectively.}
  \label{fig:foldedPatch}
\end{figure}

In this section a more complicated example is considered, where the farfield of a triple slot patch
antenna fixed on a dielectric substrate is computed. The structure is taken from the examples of
CST Microwave Studio as part of the CST Studio Suite~\cite{CST}. It is illustrated in Fig.~\ref{fig:foldedPatch}
with the defining points 1-13 of the patches given in Tab.~\ref{tab:patchsetup}. The relative permittivity
of the substrate is 2.2. The substrate and metallization thicknesses are 0.813 mm and 0.2 mm, respectively.
The antenna is excited using two discrete
voltage ports, which impose a voltage across the gaps of the antenna feed at the position of points 2 and 3.
The excitation voltage follows a Gaussian time profile with a standard deviation of 0.12 ns. The total simulation
time is 2.5 ns.

\begin{table}[bt]
  \renewcommand{\arraystretch}{1.3}
  \label{tab:patchsetup}
  \centering
  \begin{tabular}{c r r||c r r}
    \hline\hline
    Point  & $x$ / mm &  $y$ / mm & Point  & $x$ / mm &  $y$ / mm \\
    \hline
     1 & -58.5 & 0 & 8 & 37 & 32     \\
     2 & -1.7 & 0  & 9 & -37 & 34    \\ 
     3 &1.7 & 0  &   10 & 37 & 36    \\
     4 & 58.5  & 0  &11 & -37 & 38   \\
     5 & 39 & 30 &   12 & -39 & 40   \\
     6 & -1 & 32 &   13 & -58.5 & 60 \\
     7 & 1 & 32 &  & &   \\
    \hline\hline
  \end{tabular}
  \caption{Location of points 1-13 of Fig.~\ref{fig:foldedPatch} in the $x$--$y$-plane numbered from left to right and bottom up.}
\end{table}

The farfield computation involves the determination of equivalent surface current densities
on a collection surface $\Gamma$, and the subsequent solution of the Stratton-Chu integral
under the farfield assumption
\begin{equation}
  \label{eq:farfield}
  \vE_\infty(\vec{\vxn}) = \frac{ik}{4 \pi}  \,
  \int_\Gamma [ 
  \vxn \times \vec{M}(\vy) + 
  Z\vxn \times (\vxn \times \vec{J}(\vy)) 
  ] e^{ik\vxn\cdot\vy} \, \dd A,
\end{equation}
where $\vxn$ is an observation direction, $\vy$ the integration variable, $k$ the wave number
and $\vec{n}$ the inward facing unit normal.
The equivalent current densities on the collection surface are $\vec{J}(\vy) = \vec{n} \times \vec{H}(\vy)$ 
and $\vec{M}(\vy) = \vec{E}(\vy) \times \vec{n}$.
As time-domain simulations are performed a Fourier transform of the equivalent
currents involving the target frequency has to be carried out prior to solving~\eqref{eq:farfield}.

In this example, the collection surface is a box enclosing the structure at a distance of 2 mm.
Usually the mesh is constructed such that the collection surface is obtained as the union of faces of
a number of connected elements. In this case the elements, which have to be considered for
solving the farfield integral can be determined in a preprocessing step. As this advantage cannot be
exploited on adaptive meshes, we allow for placing the collection surface $\Gamma$ independently of the mesh.
The farfield integral is computed by dissecting $\Gamma$ into (mesh independent) patches and performing
a Gauss-Legendre quadrature on each patch. 
The computational domain is terminated by Silver-M\"uller radiation boundary conditions.

\begin{figure*}[t]
  \centering
  \includegraphics[width=5.5in]{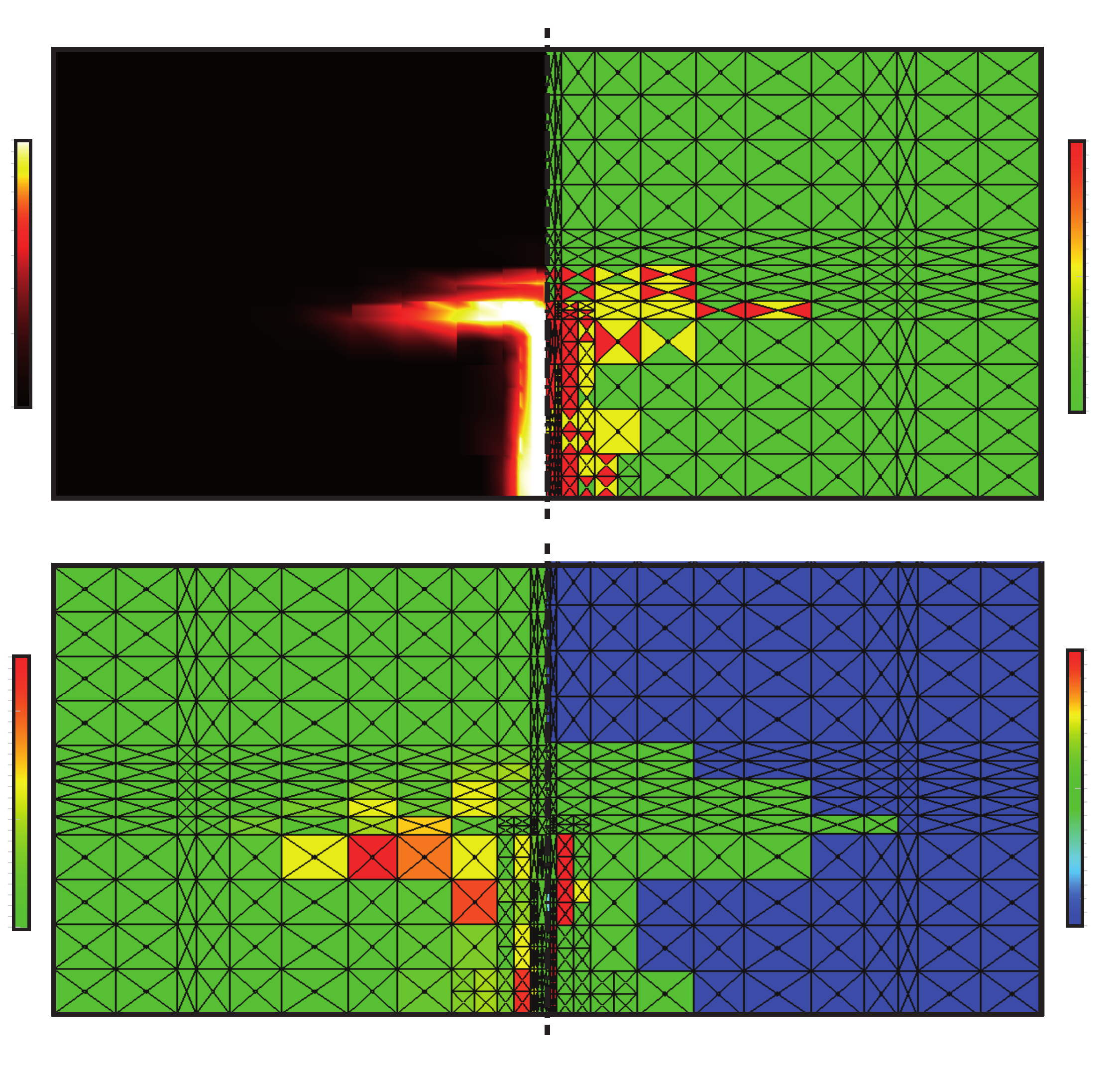}
  \put(-398,233){\text{\footnotesize 1}}
  \put(-406,326){\text{\footnotesize 10}}
  \put(-402,337){\text{\small V/m}}
  \put(-2,232){\text{\footnotesize 1}}
  \put(-2,280){\text{\footnotesize 2}}
  \put(-2,328){\text{\footnotesize 3}}
  \put(-12,337){\text{\small $P_d$}}
  \put(-409,144){\text{\footnotesize 1e-8}}
  \put(-413,93){\text{\footnotesize 5e-9}}
  \put(-413,43){\text{\footnotesize 1e-15}}
  \put(-395,157){\text{\small $\varepsilon_i$}}
  \put(-2,144){\text{\footnotesize non-refinable}}
  \put(-2,121){\text{\footnotesize refine}}
  \put(-2,96){\text{\footnotesize retain}}
  \put(-2,72){\text{\footnotesize reduce}}
  \put(-2,48){\text{\footnotesize irreducible}}
  \put(-12,157){\text{\small Marker}}
  \put(-333,192){\text{\small Electric field magnitude}}
  \put(-320,4){\text{\small Estimated error}}
  \put(-127,192){\text{\small $hp$-mesh}}
  \put(-135,4){\text{\small Adaptivity marker}}
  \caption{Snapshot at time 0.3 ns of the electric field magnitude with logarithmic color scale (top left),
    the $hp$-mesh (top right), the estimated error (bottom left) and the element marker (bottom right).
    The viewplane is located at the bottom of the substrate. A non-equidistant base mesh
    was used for capturing the edges of the patches.
    An element is marked as non-refinable/irreducible only if no more refinement/derefinement
    option ($h$ or $p$) is available,
    i.e., the given maximum/minimum $h$-level and order is met in all directions.}
  \label{fig:FieldsAdap0}
\end{figure*}

\begin{figure*}[t]
  \centering
  \includegraphics[width=5.5in]{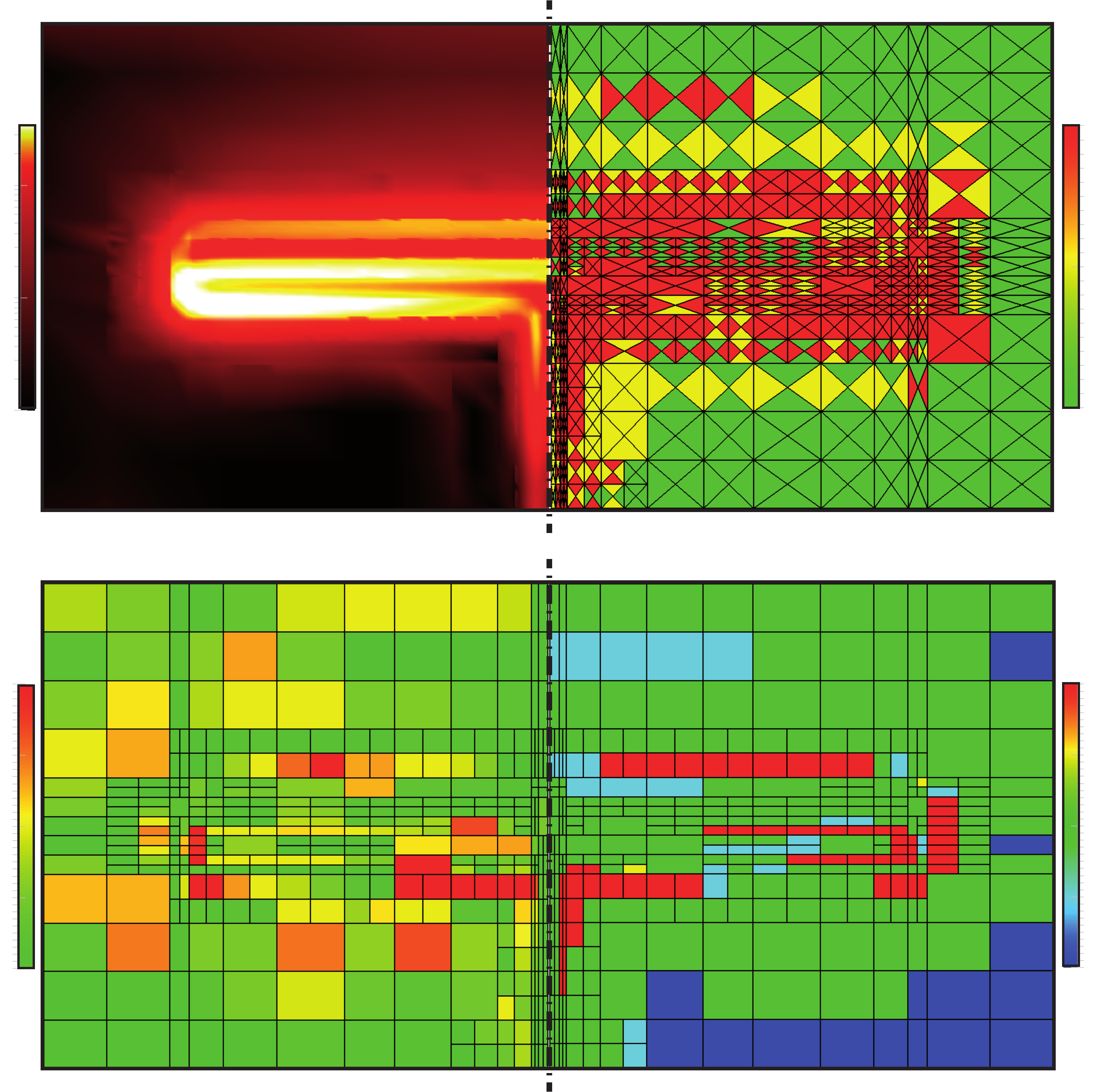}
  \put(-398,244){\text{\footnotesize 1}}
  \put(-403,284){\text{\footnotesize 10}}
  \put(-406,324){\text{\footnotesize 100}}
  \put(-406,345){\text{\footnotesize 300}}
  \put(-402,357){\text{\small V/m}}
  \put(-2,244){\text{\footnotesize 1}}
  \put(-2,297){\text{\footnotesize 2}}
  \put(-2,345){\text{\footnotesize 3}}
  \put(-12,357){\text{\small $P_d$}}
  \put(-409,144){\text{\footnotesize 5e-8}}
  \put(-413,93){\text{\footnotesize 5e-12}}
  \put(-413,43){\text{\footnotesize 5e-16}}
  \put(-395,157){\text{\small $\varepsilon_i$}}
  \put(-2,144){\text{\footnotesize non-refinable}}
  \put(-2,119){\text{\footnotesize refine}}
  \put(-2,93){\text{\footnotesize retain}}
  \put(-2,68){\text{\footnotesize reduce}}
  \put(-2,43){\text{\footnotesize irreducible}}
  \put(-12,157){\text{\small Marker}}
  \put(-333,196){\text{\small Electric field magnitude}}
  \put(-320,-3){\text{\small Estimated error}}
  \put(-127,196){\text{\small $hp$-mesh}}
  \put(-135,-3){\text{\small Adaptivity marker}}
  \caption{Snapshot at time 0.9 ns for the identical setup as in Fig.~\ref{fig:FieldsAdap0}.}
  \label{fig:FieldsAdap}
\end{figure*}

The Figures~\ref{fig:FieldsAdap0} and~\ref{fig:FieldsAdap}
show snapshots of the electric field magnitude using a logarithmic
color scale in the top left panel, the $hp$-mesh (top right), the elementwise error estimate (bottom left) and
the element markers (bottom right). The viewplane is located at the bottom of the substrate.
The physical times correspond to 0.3 ns and 0.9 ns.
Tab.~\ref{tab:hpVSfixedPatch} summarizes the number of DoF,
runtime and error estimates after the final time step for various 
adaptive and non-adaptive settings. The computing time and number of DoF
is reduced by factors of about three to five. 

\begin{figure}[tb]
  \centering
  \includegraphics[width=3in]{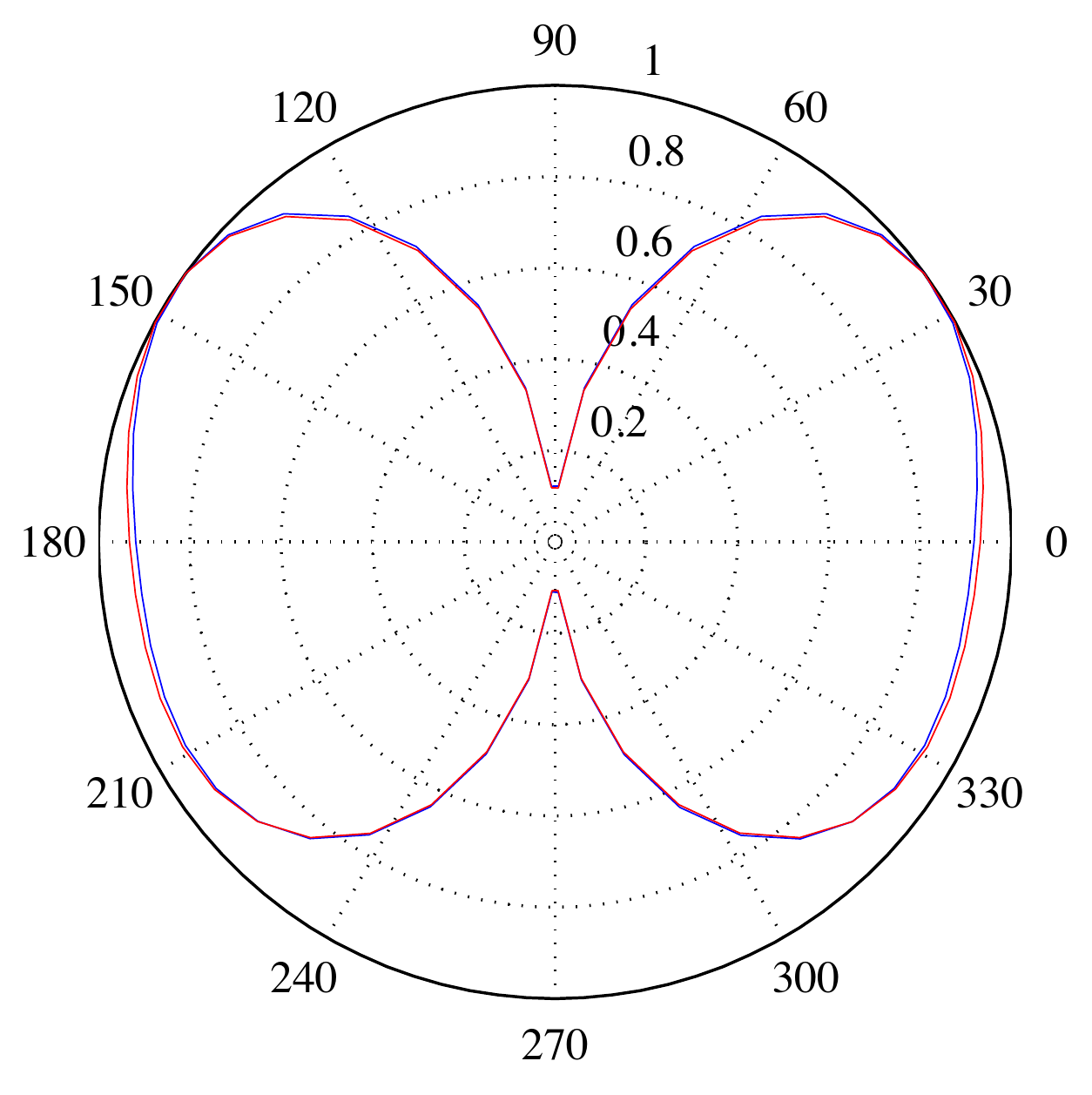}
  \includegraphics[width=3in]{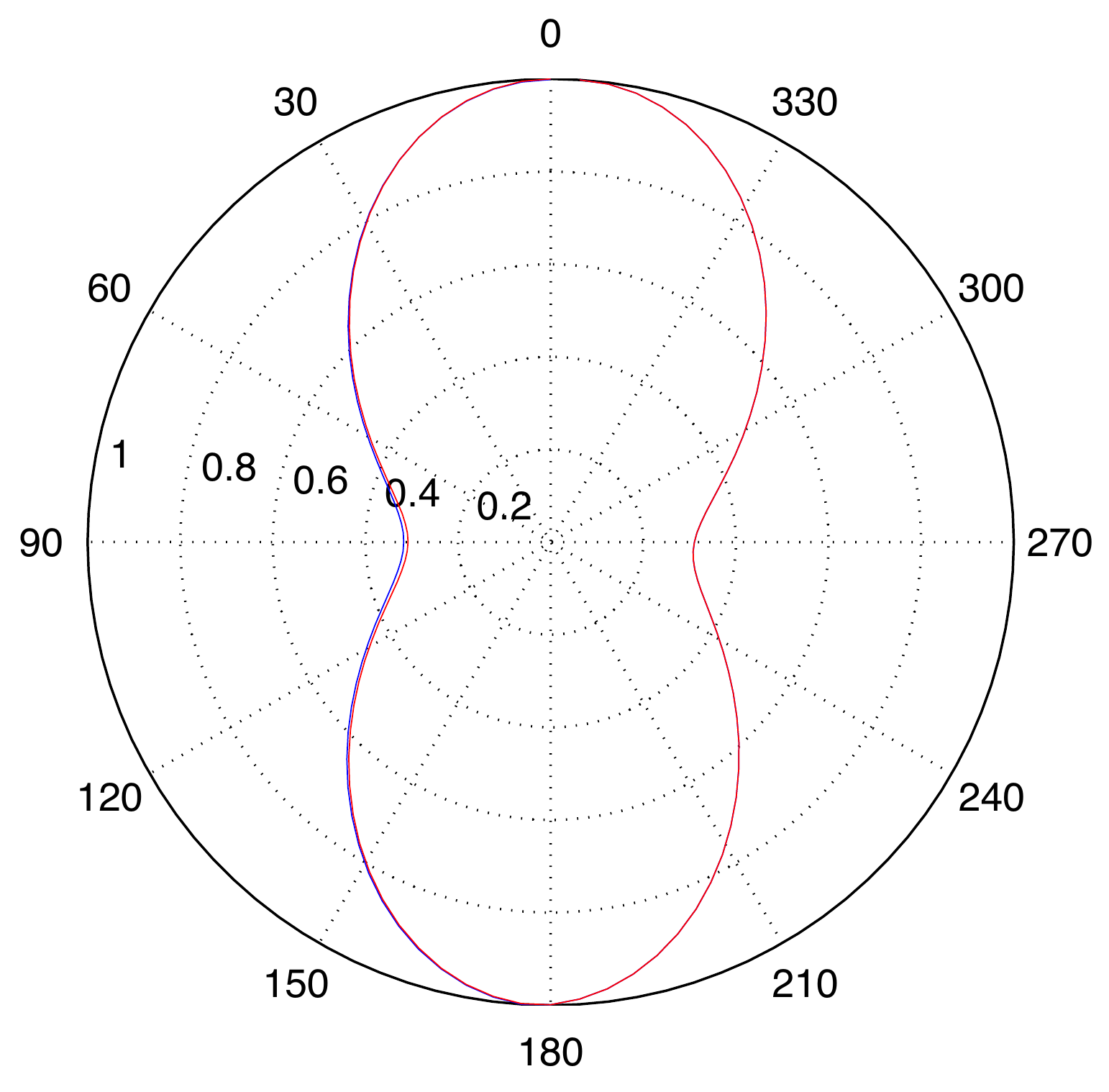}
  \caption{Normalized electric farfield at a frequency of 1.5 GHz of the triple slot patch antenna depicted in Fig.~\ref{fig:foldedPatch}. 
    The azimuth ($x$--$y$) and elevation ($z$--$x$) plane are shown in the left and right panels, respectively.
    Red curves correspond to the reference solution computed on a non-adaptive fine mesh using third order
    elements (cf.~Tab.~\ref{tab:hpVSfixedPatch} \#1), blue curves were obtained with the adaptive scheme
    and settings according to Tab.~\ref{tab:hpVSfixedPatch} \#2.}
  \label{fig:FF}
\end{figure}

The farfields computed from the reference and the setting of $\#2$ (cf.~Tab.~\ref{tab:hpVSfixedPatch}) are shown in Fig.~\ref{fig:FF}.
In this context, the approach presented in \cite{Monk:1999kd} is of interest, where the farfield error instead of the
global solution error is employed for driving mesh adaptation.

\begin{table}[bt]
  \renewcommand{\arraystretch}{1.3}
  \label{tab:hpVSfixedPatch}
  \centering
  \begin{tabular}{c r r r r r r r}
    \hline\hline
    \# & $P$ & $L$  & Elements& DoF / $10^3$ &  norm. Time & $L^2$-error / $10^{-2}$ & \texttt{TOL} / $10^{-2}$\\
    \hline
     1&{3} & --- & 11968 & 4596 & {1} & 0.89 (100 \%) & --- \\
     2&{1-3} & 0 & 11968& 574-850  & {0.15} & 1.06  (119 \%)& 0.89 \\
     3&{1-3}  & 0-1 & 3542-7446& 574- 1303 & {0.27} & 1.11 (125 \%) & 0.89\\
     4&{1-4}  & 0-1 & 3542-6454& 574- 1807 & {0.34} & 0.86  (97 \% )& 0.89\\
     5&{2} & --- & 11968& 1939 & {0.21} & 1.56  (175 \%)& --- \\
     6&{1} & --- & 11968& 574 & {0.03} & 4.60  (517 \%)& --- \\
    \hline\hline
  \end{tabular}
  \caption{Performance of simulations of example~\ref{sec:folded-patch-antenna} using fixed and adaptive meshes.
    The results given in the first row (\#1) obtained on a static mesh of third order elements
    are taken as a reference. \#2-\#4 were obtained on various adaptive meshes.
    In particular, \#2 was obtained using the same topological mesh as \#1 using pure
    $p$-refinement with orders between one and three. Results \#3 and \#4 were obtained using a coarser
    root mesh with a maximum of one level of $h$-refinement. All adaptive simulations use
    as error tolerance the error of the reference, i.e.~$0.89\cdot10^{-2}$. 
    This tolerance is not met in the cases \#2 and \#3, which is expected as
    the local resolution in every element is less or equal compared to the reference.
    In \#4 a maximum order of 4 is permitted and the error tolerance is met.
    Results \#5 and \#6 employ the same topological mesh as the reference but uniform element
    orders of two and one, respectively. The computing time of \#5 is comparable to the adaptive solutions,
    but it has a much larger error.}
\end{table}



\section{Conclusion and Outlook}
\label{sec:conclusion}

A scheme for performing time-domain simulations with the DG method on
anisotropically refined dynamic $hp$-meshes in three-dimensional space was proposed. The adaptation is driven by
the local solution error and guided by a novel variant of the concept of reference
solutions. It drastically reduces the computational costs associated with error and regularity estimation
allowing for the first time to perform fully automatic $hp$-adaptation for three-dimensional
transient problems, where a given error tolerance is respected throughout the simulation. 
This was achieved by interchanging the role of the reference mesh
and the solution mesh in the construction of the error estimate. While this comes at the cost of losing
some sharpness of the estimate, it largely increases the practical applicability of the approach.
The computation of the proposed error estimate is highly efficient as it is free of quadratures.
Code profiling showed that for the presented examples the computational time consumed for all adaptivity related tasks
was around 15~\% of the total computing time. 

The attainable savings in terms of computing time and memory consumption
using dynamical $hp$-meshes strongly depend on the application. They
roughly scale with the multi-scale character of the problem at hand
and can reach factors above one hundred~\cite{Schnepp:2012JCAM}.
Here, two examples were shown, where computation times were reduced up to a factor of 20.
Savings are particularly large with respect to implementations employing isotropic approximation
orders only.

The implementation is currently restricted to orthogonal hexahedral meshes, which imposes 
limitations for the modeling of arbitrary structures. However, the proposed
adaptation algorithm is independent of the actual element shape and can 
be applied on non-orthogonal curvilinear hexahedral meshes as well. This is the subject of ongoing work.



\end{document}